\documentclass[twocolumn,superscriptaddress,longbibliography,aps,prb,preprintnumbers]{revtex4-2}

\usepackage[normalem]{ulem}
\usepackage{graphicx}
\usepackage{bm}
\usepackage{color}
\usepackage{epsfig}
\usepackage{amsmath}
\usepackage{amssymb}
\usepackage{wrapfig}
\usepackage{natbib}

\usepackage{hyperref}
\hypersetup{
    colorlinks=true,      
    urlcolor=blue,
    citecolor=blue,
    linkcolor=blue
}
\begin{document}
\title{Anomalous Hall and Nernst effect switching via staggered rotation in a kagome antiferromagnetic semimetal}

\author{Subhadip Pradhan}
\affiliation{National Institute of Science Education and Research, Jatni, Odisha 752050, India}
\affiliation{Homi Bhabha National Institute, Training School Complex, Anushakti Nagar, Mumbai 400094, India}
\author{Kartik Samanta}
\affiliation{Department of Physics and Astronomy $\&$ Nebraska Center for Materials and Nanoscience, University of Nebraska, Lincoln, NE 68588, USA}
\author{Ashis K. Nandy}
\email{aknandy@niser.ac.in}
\affiliation{National Institute of Science Education and Research, Jatni, Odisha 752050, India}
\affiliation{Homi Bhabha National Institute, Training School Complex, Anushakti Nagar, Mumbai 400094, India}
\begin{abstract}
The intricate interplay between magnetism and the topology of electronic structures provides a rich avenue for tailoring materials with unique and potent anomalous transport properties. In this paper, we present a strategy for inducing robust Berry curvature and anomalous transverse conductivity in noncollinear antiferromagnets through an unconventional approach termed ``small \textit{staggered rotation} of spin". Considering noncollinear Mn$_3$Sn, we demonstrate that the positive vector chirality antiferromagnetic configuration, typically associated with a vanishing anomalous Hall effect and Nernst effect, can be manipulated to exhibit finite anomalous Hall conductivity (AHC) and anomalous Nernst conductivity (ANC) through \textit{staggered rotation}. Furthermore, we illustrate that the value and sign of both the AHC and ANC can be tuned through \textit{staggered rotation}. This tuning is intricately influenced by the spin-orbit coupling (SOC) induced gapped nodal line, revealing the critical role of electronic structure modifications in achieving precise control over transport properties.

\end{abstract}

\maketitle


\section{\label{sec:level1}INTRODUCTION}
In recent years, materials exhibiting strong anomalous transport properties have captured significant attention within the condensed matter community. The intimate connection between a material's transport characteristics and its band topology has spurred widespread interest in the exploration of the latter. Among the most pivotal band topologies that have garnered attention, topological semimetals, particularly nodal line semimetals~\cite{Fang_2016,PhysRevB.92.045108,PhysRevB.108.115122} and Weyl semimetals~\cite{Yan_2017,RevModPhys.90.015001,Yang_2016,PhysRevB.83.205101}, hold a central position. In the case of Weyl semimetals, the presence of a Weyl-type band crossing leads to a notably high Berry curvature, consequently generating a heightened intrinsic anomalous Hall conductivity (AHC)~\cite{RevModPhys.90.015001}. Conversely, nodal line semimetals display distinctive features due to spin-orbit coupling (SOC) induced gaps, leading to a substantial generation of Berry curvature and the consequent manifestation of anomalous transport properties~\cite{article,PhysRevB.99.165117}. The stability of a topological nodal line semimetal is intricately linked to specific crystalline and mirror symmetries, and the disruption of these symmetries through SOC induces a significant intrinsic Berry curvature in the system~\cite{Fang_2016}. The interplay between band topology and symmetry-breaking mechanisms shapes the unique transport properties of these materials, making them interesting subjects for further investigation.\\
While most spintronic devices have traditionally relied on ferromagnetic materials, recent focus has shifted towards antiferromagnetic materials~\cite{RevModPhys.90.015005,Salemi_2019,Jungwirth_2016}. The resurgence of interest in noncollinear antiferromagnets (AFMs) Mn$_3$X (X = Ir, Pt)~\cite{PhysRevB.92.144426,PhysRevLett.112.017205} and Mn$_3$Y (Y = Sn, Ge, Ga)~\cite{Kubler_2017,Yang_2017,WANG2022100878,Park_2018,Singh_2024,PhysRevB.110.094432} is driven by the discovery of novel phenomena, such as the substantial anomalous Hall effect (AHE)~\cite{K_bler_2014,nakatsuji2015large,doi:10.1126/sciadv.1501870, PhysRevB.95.075128}, anomalous Nernst effect (ANE)~\cite{Ikhlas_2017,PhysRevLett.119.056601}, spin Hall magnetoresistance~\cite{PhysRevB.108.144435}, tunneling magnetoresistance~\cite{PhysRevLett.128.197201} and the spin Seebeck effect~\cite{10.1063/5.0045627}. Those remarkable observations in these material systems render them a captivating focus for exploring the nuanced connections between topology, electron transport, and magnetism.\\
In a non-collinear AFM, the non-collinear AFM order is characterized by the vector chirality $\kappa$, defined as~\cite{Kawamura_2001,Pradhan_2023}
\begin{equation}
\kappa = \frac{2}{3\sqrt{3}} \sum_{<ij>}[{\hat{n}_i }\times {\hat{n}_j}]_{z}
\label{VC}.
\end{equation}
For a unit magnetic moment, $\kappa$ can take on values of either $+1$ or $-1$. The configuration with $\kappa = +1$ is referred to as the ``direct" AFM configuration, while $\kappa = -1$ corresponds to the ``inverse" AFM configuration. In the Mn$_3$Y (Y = Sn, Ge) system, the ground state exhibits an ``inverse" AFM configuration with $\kappa = -1$. Having a small energy difference on the order of a few meV between the $\kappa = +1$ and $\kappa = -1$ AFM configurations, our previous study~\cite{Pradhan_2023} demonstrated that a transition from the $\kappa = +1$ to the $\kappa = -1$ AFM configuration can be achieved through a \textit{staggered rotation} of any two spins while keeping the third one unaltered in the unit cell. Both experimental and theoretical studies have highlighted a remarkably high anomalous transport property in the ``inverse" AFM phase of Mn$_3$Sn, attributed to the presence of a Weyl point in the band structure~\cite{Yang_2017,Kubler_2017,chen2021anomalous}.
\begin{figure*}
    \centering
    \includegraphics[width=\textwidth]{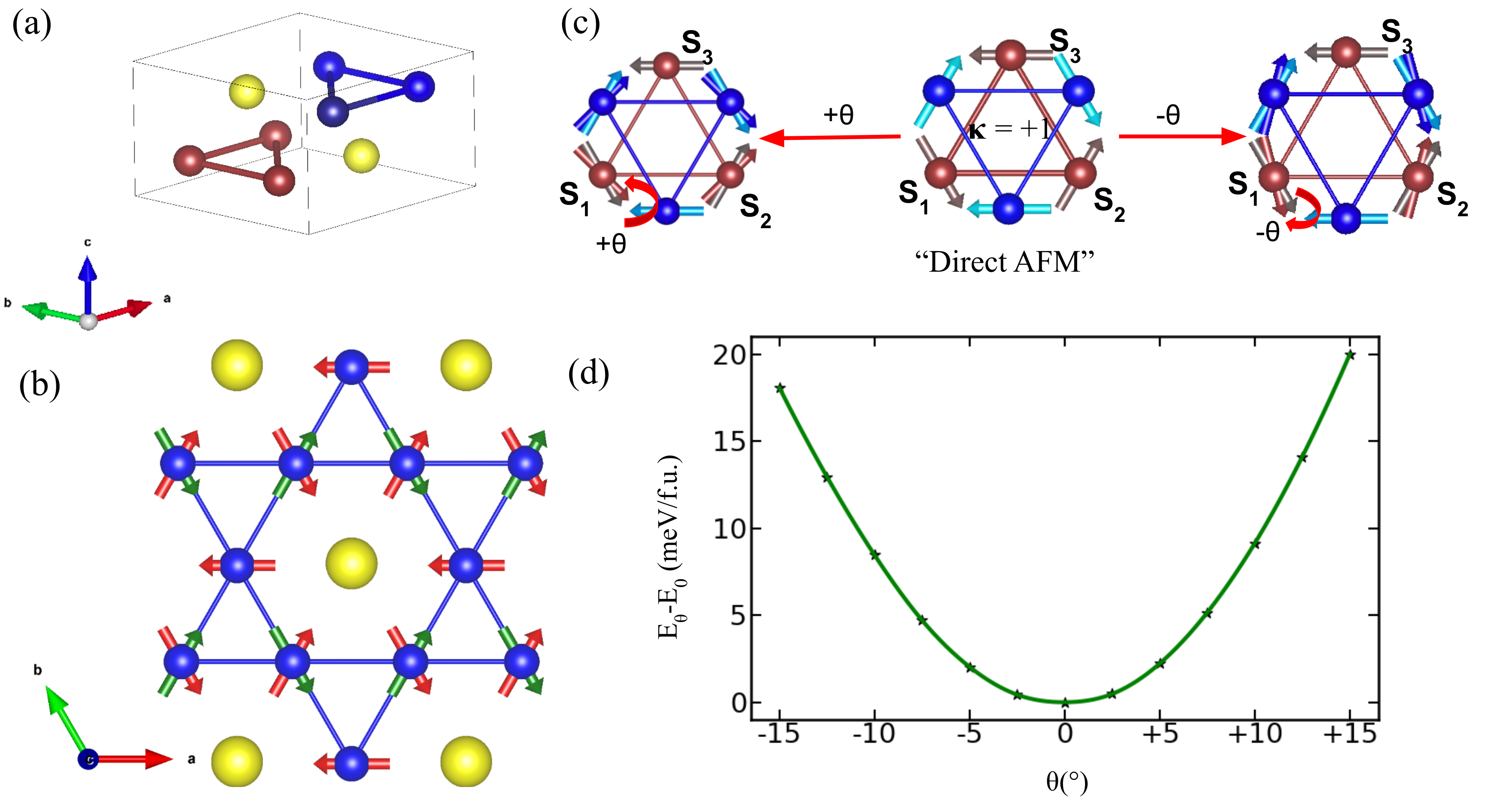}
    \caption{{\bf (a)} Bulk crystal structure of Mn$_3$Sn. The maroon and blue balls represent the magnetic Mn atoms in the two parallel kagome layers ($z = \frac{1}{4} $ and $z= \frac{3}{4}$ planes), and the Sn atoms are represented by the yellow balls. {\bf (b)} Noncollinear chiral AFM textures are illustrated in a single kagome layer. The ``direct" AFM state, with vector chirality $\kappa = +1$, is represented by red arrows, while the ``inverse" AFM state, with vector chirality $\kappa = -1$, is indicated by green arrows. {\bf (c)} An illustration of \textit{staggered rotation} between two spins, \( S_1 \) and \( S_2 \), originating from the ``direct" AFM configuration. The two distinct directions of \textit{staggered rotation} are represented as `\( +\theta \)' and `\( -\theta \)'. {\bf (d)} The calculated energy differences for different magnetic configurations are shown as a function of the \textit{staggered rotation} angle \( \theta \), with respect to the ``direct" AFM configuration.}
    \label{fig1}
\end{figure*}
In contrast, the ``direct" AFM phase lacks the symmetry required to induce a finite AHC.  Previous experimental studies have shown that this ``direct" AFM phase is reliable under pressure~\cite{PhysRevResearch.2.043366} or under different temperature gradients~\cite{temperature}. Considerable investigation has been carried out, both theoretically and experimentally, on the inverse triangular phase in Mn$_3$Sn.  However, our current study focuses primarily on the ``direct" AFM configuration of Mn$_3$Sn, revealing that a giant AHC can be generated by introducing a small \textit{staggered rotation} within this specific AFM arrangement. This type of \textit{staggered rotation} of spins with small magnitudes can be experimentally observed through the application of strain~\cite{Pizzo} or an external magnetic field~\cite{Miwa}. Under an external magnetic field, the Zeeman effect favors the alignment of all spins with the magnetic field. The interaction between one spin and its immediate neighbors promotes both clockwise and anticlockwise twists, thereby fixing the orientation of the third spin in the noncollinear AFMs~\cite{PhysRevB.106.L020402}. We also demonstrate an almost linear tuning behavior of the AHC and anomalous Nernst conductivity (ANC) as a function of the staggered rotation angle. Moreover, we elucidate how the sign of the AHC and ANC is influenced by the direction of the \textit{staggered rotation} angle. We have provided a comprehensive analysis, including ab initio calculations and symmetry considerations, to discuss the mechanisms governing the tuning of AHC and ANC.
\section{\label{sec:level2}METHOD}
The \textit{ab-initio} electronic structure calculations are performed using two different density functional theory (DFT) codes: the full-potential linearized augmented plane wave based code FLEUR~\cite{fleur} and the pseudopotential based code VASP~\cite{vasp}. We ensure the consistency of our results between the two codes by comparing total energy, band structure, and density of state calculations. The experimental structure of Mn$_3$Sn is utilized for all our calculations~\cite{PhysRevResearch.2.043366}. We employ an $8 \times 8 \times 9$ $\Gamma$-centered k-point grid and a plane-wave cutoff of 500 eV for self-consistent calculations in VASP code. The Perdew-Burke-Ernzerhof (PBE)~\cite{pbe} exchange-correlation functional and the projector-augmented-wave (PAW)~\cite{paw2} method are consistently applied throughout the calculations. In our pseudopotential based calculations, we utilize the ``I-CONSTRAINED-M'' parameter to constrain the direction of magnetic moments, allowing for the exploration of various AFM orders, \textit{staggered rotations}, and spin canting.\\
We employ a plane-wave cutoff of $k_{\text{max}} = 4.2$ a.u.$^{-1}$ to expand the linearized augmented plane wave (LAPW) basis functions in our self-consistent calculations using the FLEUR code~\cite{fleur}. A Monkhorst-Pack~\cite{monk} k-mesh of $8 \times 8 \times 9$ is used across the Brillouin zone to obtain converged charge densities. For these calculations, we utilize the Vosko-Wilk-Nusair (VWN) exchange-correlation functional~\cite{1980CaJPh..58.1200V} within the local density approximation (LDA). The plane-wave cutoffs for the potential and exchange-correlation potential are set to $14.0$ and $12.0$ a.u.$^{-1}$, respectively. SOC is incorporated self-consistently in our calculations.\\
To analyze the electronic band structures, incorporating SOC, we have employed maximally-localized Wannier functions (MLWFs)~\cite{wan2,wan4} alongside all-electron full potential LAPW methods, implemented in the FLEUR code \cite{fleur}. We construct a tight-binding Hamiltonian using MLWFs derived from Mn-d and Sn-p states, accurately representing the system's spectrum across a wide energy range centered around the Fermi energy. We repeated the Wannierization process with various inner and outer energy windows until the resulting spectrum accurately reproduced the system's spectrum within the desired energy window.
To evaluate the intrinsic contribution of AHC, we employ the Wannier interpolation technique \cite{wan1} available in the FLEUR code~\cite{Freimuth-2008}. We compute the Berry curvatures using the Wannier90 tool~\cite{wan2, wan4}, utilizing a well-constructed tight-binding model based on MLWFs. The linear response Kubo formula \cite{Yao-2004} is applied to compute the AHC. We have calculated the ANC using the method proposed by Xiao et al~\cite{PhysRevLett.97.026603,RevModPhys.82.1959}.
We have employed a $300\times300\times300$ k-point mesh to calculate the AHC and ANC. For the ANC calculation, we have considered a temperature of T = 300 K. The WannierBerri Python code~\cite{Tsirkin_2021} is used to map the Berry curvature onto the band structure.
\section{\label{sec:level3}MAGNETIC STRUCTURE OF MN$_3$SN, SYMMETRY ANALYSIS AND TRANSPORT PROPERTY AS A FUNCTION OF STAGGERED ROTATION}
\subsection{Magnetic structure of ``direct" AFM Mn$_3$Sn and staggered rotation of spin} Fig.~\ref{fig1}(a) illustrates the layered hexagonal lattice structure of bulk Mn$_3$Sn, which crystallizes in the space group \( P6_3/mmc \) (No. 22). Within this structure, Mn atoms form kagome layers in the ab plane, and these kagome planes are stacked along the c axis. Sn atoms are positioned at the center of the hexagon formed by the Mn atoms, indicating that the Sn atoms reside in the same plane as the Mn atoms. It possesses experimental lattice parameters with $a = b = 5.65 $  \AA~  and $c = 4.522$ \AA. The magnetic structure of Mn\(_3\)Sn exhibits notable complexity across various temperature ranges. Below the Néel temperature (\(T_N = 420 \, \text{K}\)), the system enters an AFM phase. In this phase, the magnetic moments of the Mn atoms align within the \(ab\)-plane, forming a \(120^\circ\) noncollinear configuration. Within this non-collinear AFM phase, two competing magnetic configurations, defined by vector chiralities \(\kappa = +1\) and \(\kappa = -1\), are possible. These distinct configurations are depicted in Fig.~\ref{fig1}(b), where the spins, represented in different colors, highlight the two AFM states. The transition between these chiral AFM states involves a \(120^\circ\) staggered rotation of two spins in the unit cell, while the third spin remains unchanged~\cite{Pradhan_2023}. \\
Here, we will concentrate only on the ``direct" AFM phase. A detailed schematic of the \textit{staggered rotation} is presented in Fig.~\ref{fig1}(c), illustrating two distinct  \textit{staggered rotation} senses, labeled as \(+\theta\) (positive rotation) and \(-\theta\) (negative rotation), relative to the ``direct” AFM configuration at \(\theta = 0\). For the \(+\theta\) case, with spin \(\mathbf{S}_3\) fixed, \(\mathbf{S}_1\) rotates counterclockwise, while \(\mathbf{S}_2\) rotates clockwise by the same angle \(\theta\). Conversely, in the \(-\theta\) configuration, \(\mathbf{S}_1\) rotates clockwise, and \(\mathbf{S}_2\) counterclockwise, maintaining the same angular displacement \(\theta\). The inclusion of SOC has a significant impact on the energetics of these staggered configurations. Configurations with a finite staggered rotation exhibit higher energy compared to the ``direct” AFM state. As shown in Fig.~\ref{fig1}(d), the energy landscape depends on the rotation direction of spins \(\mathbf{S}_1\) and \(\mathbf{S}_2\). Interestingly, the energy variation is not symmetric about \(\theta = 0\), indicating that the system's energy is influenced by the directionality of the staggered rotation. This asymmetry in the total energy may arises due to antisymmetric exchange interactions, a hallmark of the Dzyaloshinskii-Moriya interaction (DMI). This suggests that different staggered rotations could give rise to distinct electronic properties, underscoring the intricate interplay between spin structure and electronic behavior in Mn\(_3\)Sn.
\begin{figure}[ht!]
    \centering
     \includegraphics[width=0.48\textwidth]{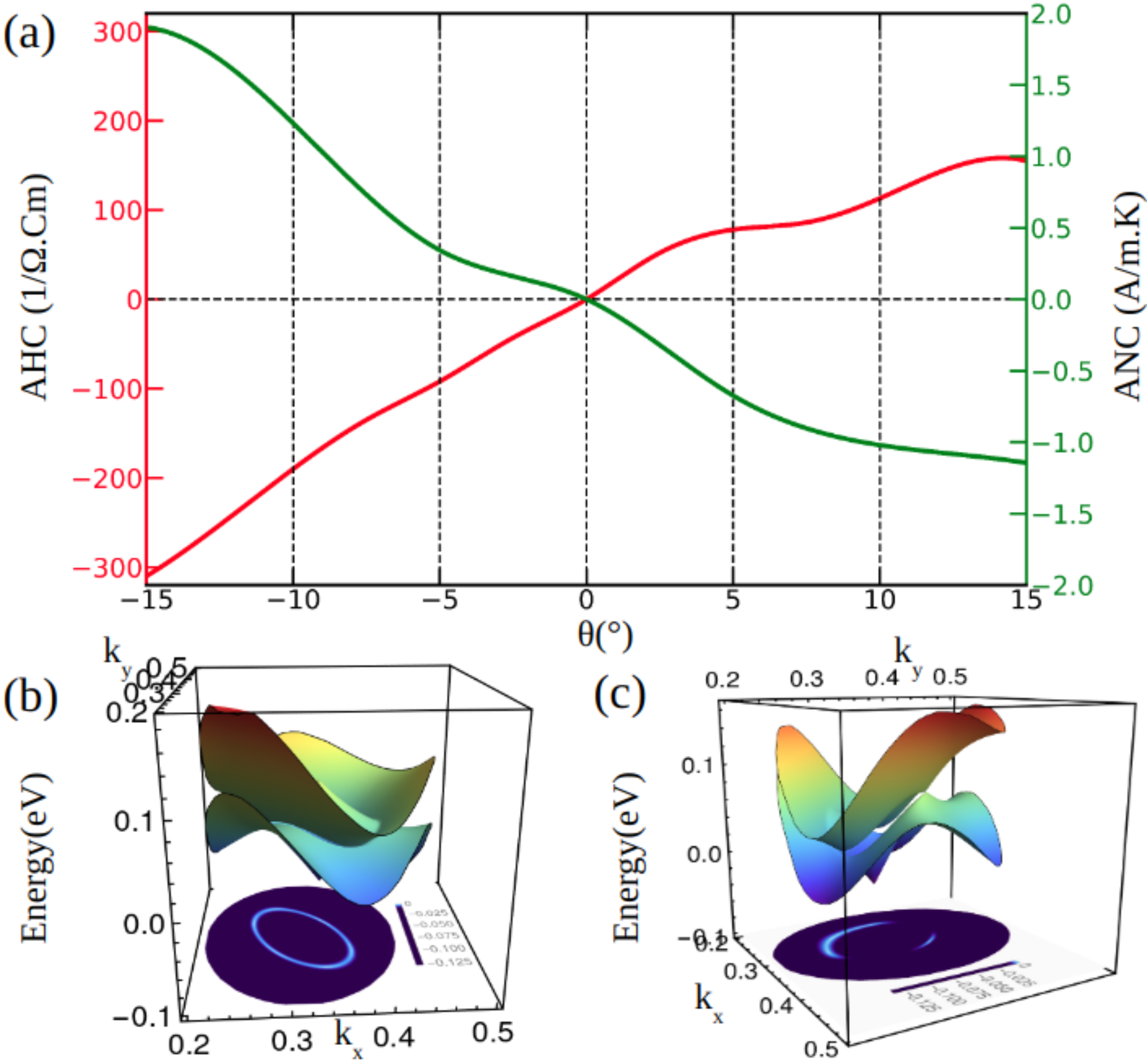}
     \caption{ {\bf (a)} \textit{Staggered rotation} angle dependent anomalous Hall conductivity (AHC) and anomalous Nernst conductivity (ANC) for ``direct" AFM phase of Mn$_3$Sn. Starting from the initial ``direct" AFM state of Mn$_3$Sn, finite AHC (red) and the ANC (green) emerge when a finite \textit{staggered rotation} angle is introduced. The anomalous transport coefficients vary between `+' and `-' depending on the staggered rotation angle. {\bf (b)} and {\bf (c)} The bulk 3D band structure in the $k_z = 0$ plane with spin-orbit coupling (SOC) for the ``direct" AFM configuration and for a staggered rotation of `$-5^{o}$' from that configuration. The figures illustrate that with a finite \textit{staggered rotation}, a nodal line gap opens in the bulk band structure.} 
     \label{fig:ahc_tuning}
\end{figure}
\subsection{Symmetry analysis for AHC}
The Anomalous Hall Conductivity (AHC) tensor can be written as~\cite{RevModPhys.82.1539},
\begin{equation}
    \sigma_{ij}^{AH} =-\epsilon_{ijl}~\frac{e^2}{\hbar}\sum_n \int \frac{d\mathbf{k}}{(2\pi)^d} f(\epsilon_n(\mathbf{k})) \Omega_{n}^{\alpha}(\mathbf{k})
\end{equation}
Where $\epsilon_{ijl}$ is the antisymmetric tensor, where $i, j = x, y, z$ and $n$ is the band index. $f_n(\mathbf{k})$ is the Fermi-Dirac distribution.
The Berry curvature corresponding to state $n(\mathbf{k})$ is defined as,
\begin{equation}
    \bm{\Omega}_{n}(\mathbf{k}) = \bm{\nabla}_{\mathbf{k}} \times \mathbf{a}_{n}(\mathbf{k})
\end{equation}
The presence of a finite AHC component, denoted by \( \sigma_{\alpha} \)($\sigma_{ij}$), in a magnetic system depends on the symmetry of the Berry curvature in momentum space. If the Berry curvature satisfies the condition \( \Omega^{\alpha}(R\mathbf{k}) = -\Omega^{\alpha}(\mathbf{k}) \) under the magnetic symmetry operation R, the corresponding AHC component \( \sigma_{\alpha} \) must be zero, as the Berry curvature contributions at \( \mathbf{k} \) and \( R\mathbf{k} \) cancel out during Brillouin Zone (BZ) integration~\cite{PhysRevB.104.125145,PhysRevB.95.094406}. In a magnetic system with an \(n\)-fold rotation symmetry, the Berry curvature cancels out during BZ integration. For example, in the \(d_{6h}\) point group with a three-fold rotational symmetry about the z-axis, the components of \( \mathbf{\sigma} \) perpendicular to the z-axis, i.e., \( \sigma_{x} \) and \( \sigma_{y} \), are zero.\\
\begin{figure*}[ht!]
    \centering     
    \includegraphics[width=\textwidth]{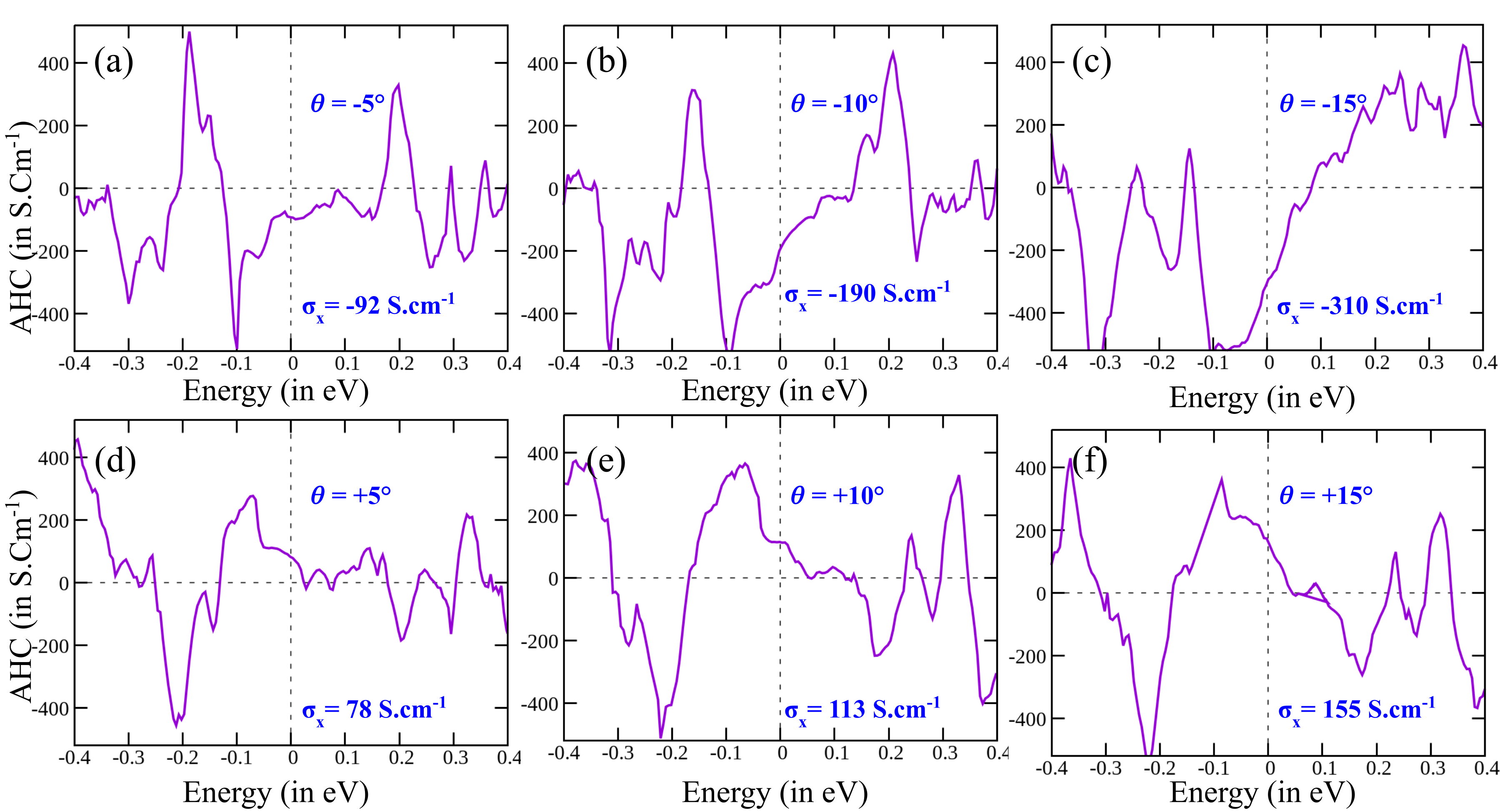}
     \caption{{\bf(a)}-{\bf(f)} Energy dependent anomalous Hall conductivity (AHC) for \textit{staggered rotation} angles of \( -5^\circ \), \( -10^\circ \), \( -15^\circ \), \( +5^\circ \), \( +10^\circ \), and \( +15^\circ \). } 
     \label{fig:AHC_Scan}
\end{figure*}
\begin{figure*}[ht!]
\centering
\includegraphics[width=\textwidth]{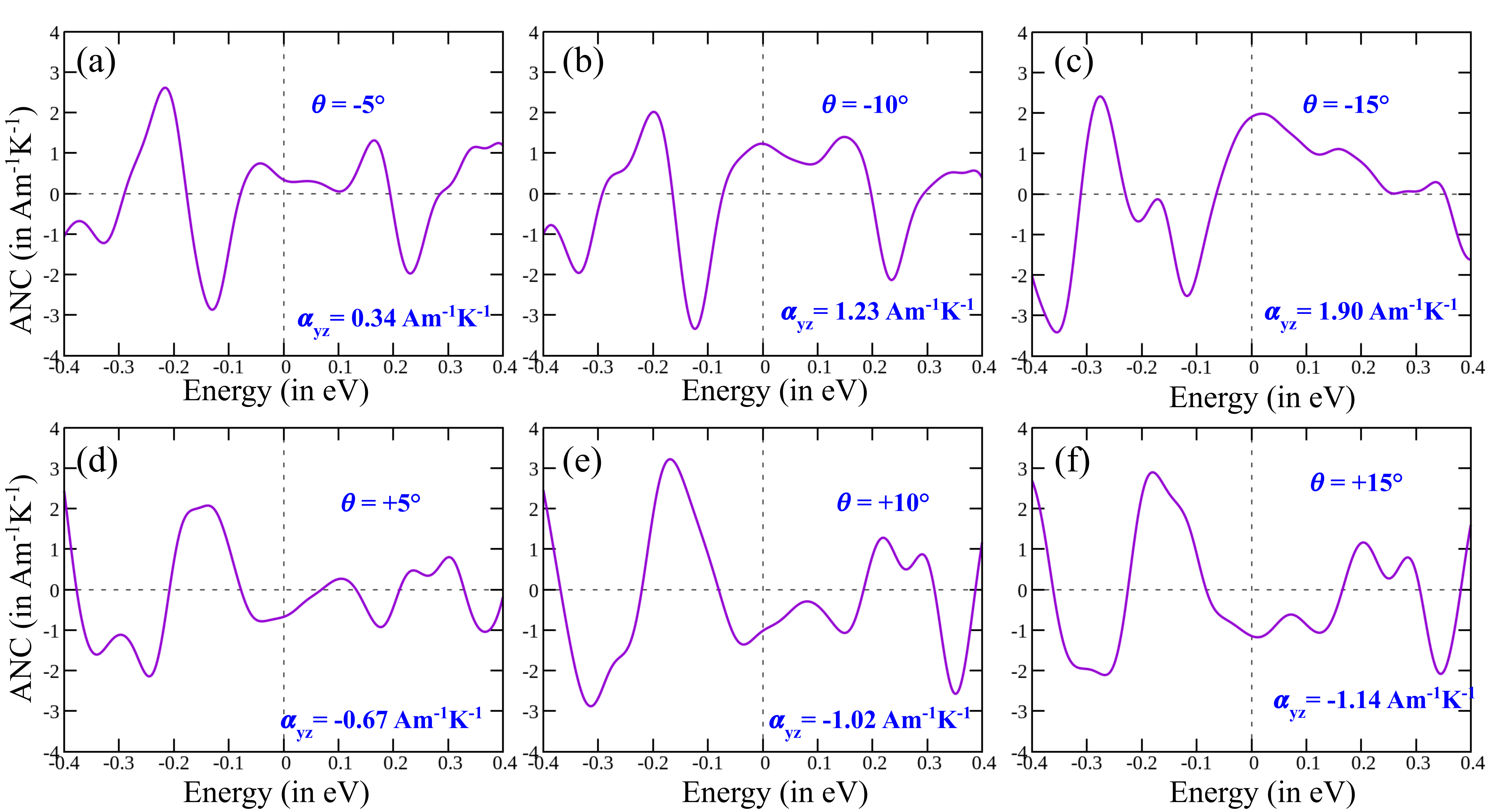}
     \caption{{\bf (a)}-{\bf (f)} Energy dependent anomalous Nernst conductivity (ANC) for \textit{staggered rotation} angles of \( -5^\circ \), \( -10^\circ \), \( -15^\circ \), \( +5^\circ \), \( +10^\circ \), and \( +15^\circ \).} 
     \label{fig:ANC_Scan}
\end{figure*}
\begin{figure*}[ht!]
    \centering
     \includegraphics[width=\textwidth]{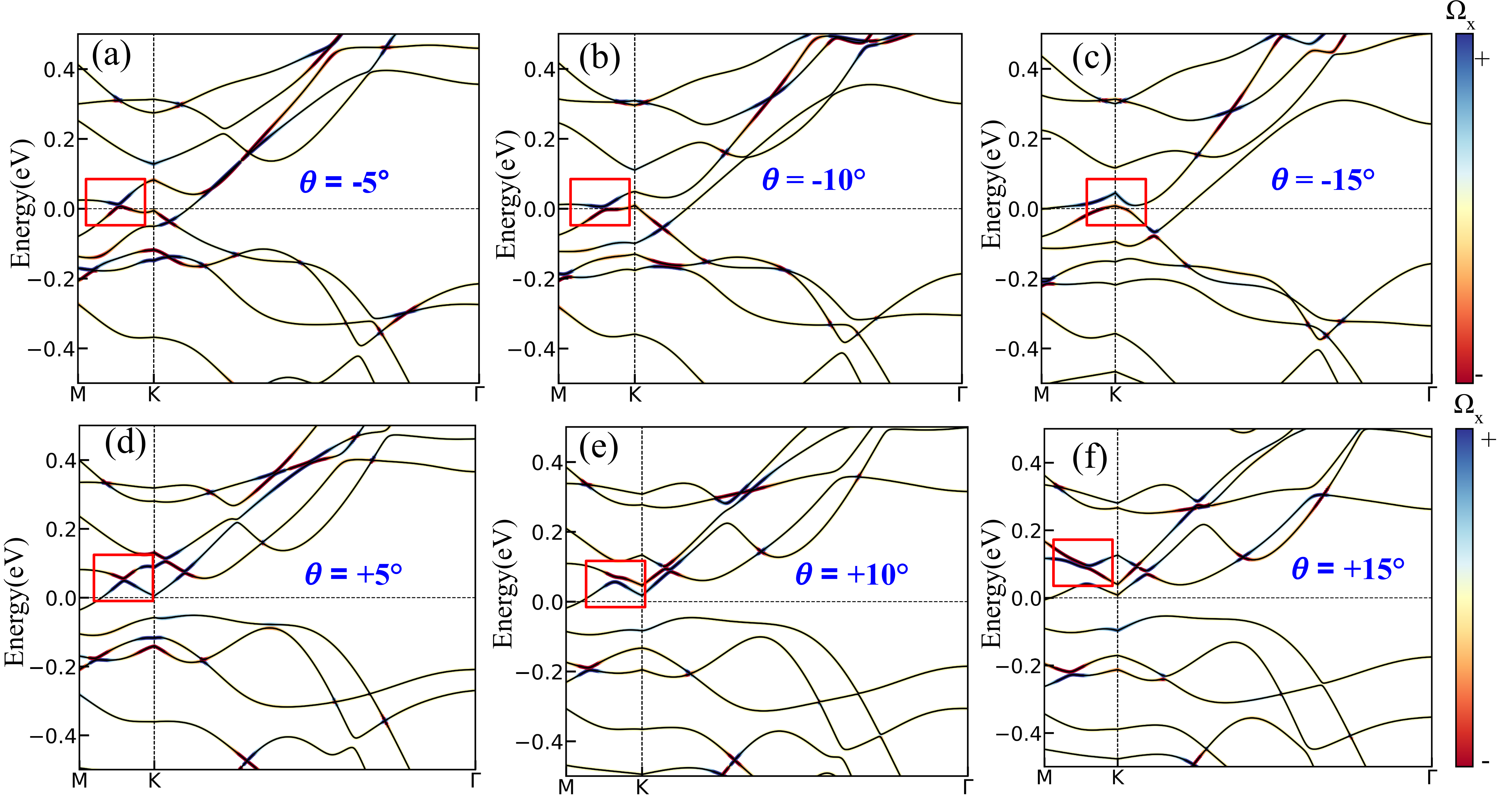}
     \caption{Band structures including SOC are shown in panels (a)-(f), incorporating the momentum-dependent Berry curvature for \textit{staggered rotation} angles of $-5^\circ$, $-10^\circ$, $-15^\circ$, $+5^\circ$, $+10^\circ$, and $+15^\circ$, respectively. The color bars on the right indicate the Berry curvature's strength, with positive values represented in blue and negative values in red. The red box highlights the two bands near the Fermi energy, showing their Berry curvature contributions.} 
     \label{fig:berry_band}
\end{figure*}
\begin{figure*}[ht!]
    \centering
     \includegraphics[width=\textwidth]{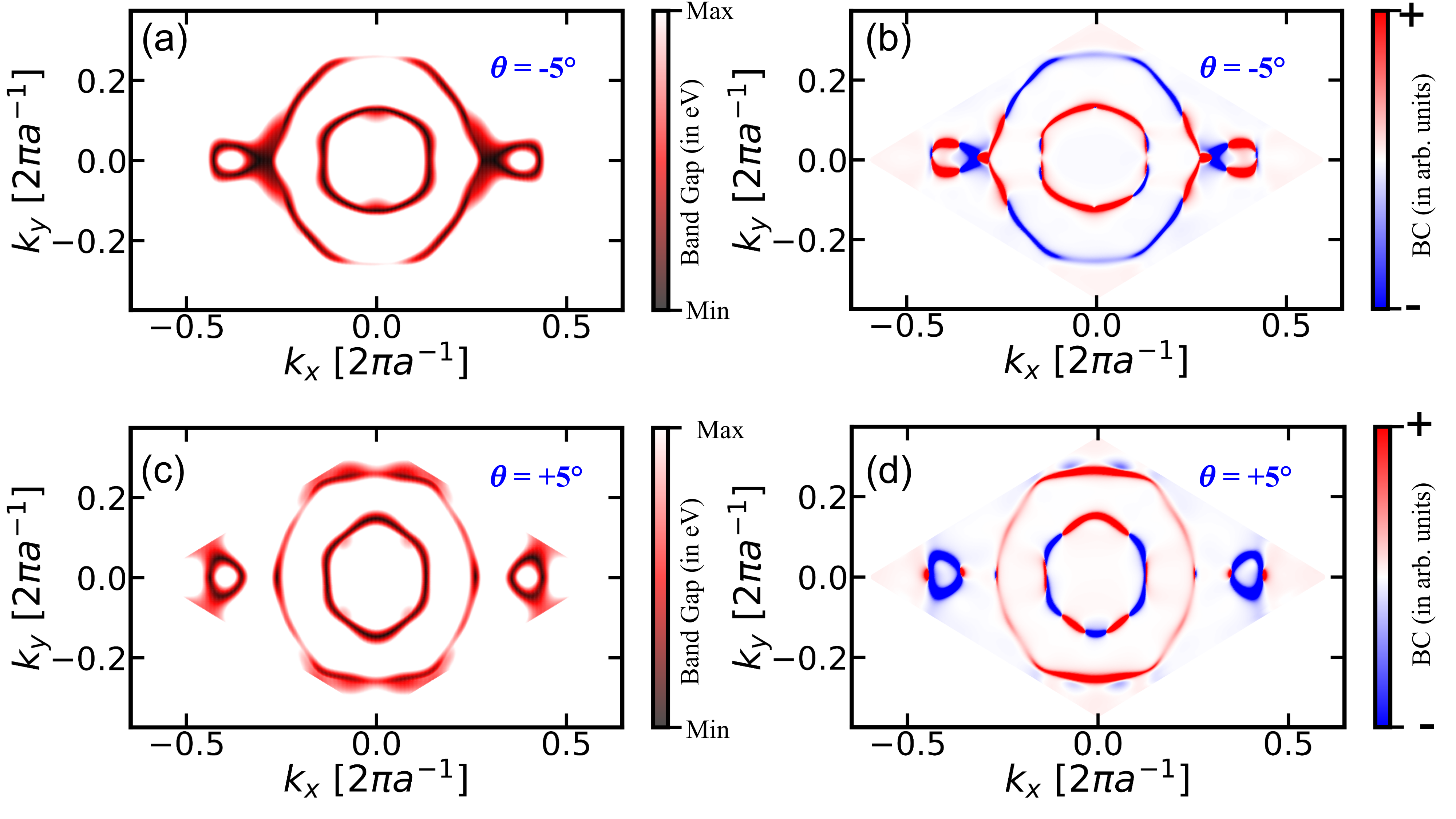}
     \caption{{\bf(a)}-{\bf(d)} The band gap between two bands near the Fermi level in the $k_z = 0$ plane is shown for (a) $\theta = -5^\circ$ and (c) $\theta = +5^\circ$. In both configurations, gapped nodal line features are visible. Panels (b) and (d) show the corresponding Berry curvature distribution in the $k_x$-$k_y$ plane for $\theta = -5^\circ$ and $\theta = +5^\circ$, respectively, where significant Berry curvature contributions are observed along the gapped nodal lines. When \( \theta = -5^\circ \), the Berry curvature contribution is maximally negative, whereas at \( \theta = +5^\circ \), it is maximally positive.} 
     \label{fig:BC_2D}
\end{figure*}
\begin{figure}[ht!]
    \centering
     \includegraphics[width=0.48\textwidth]{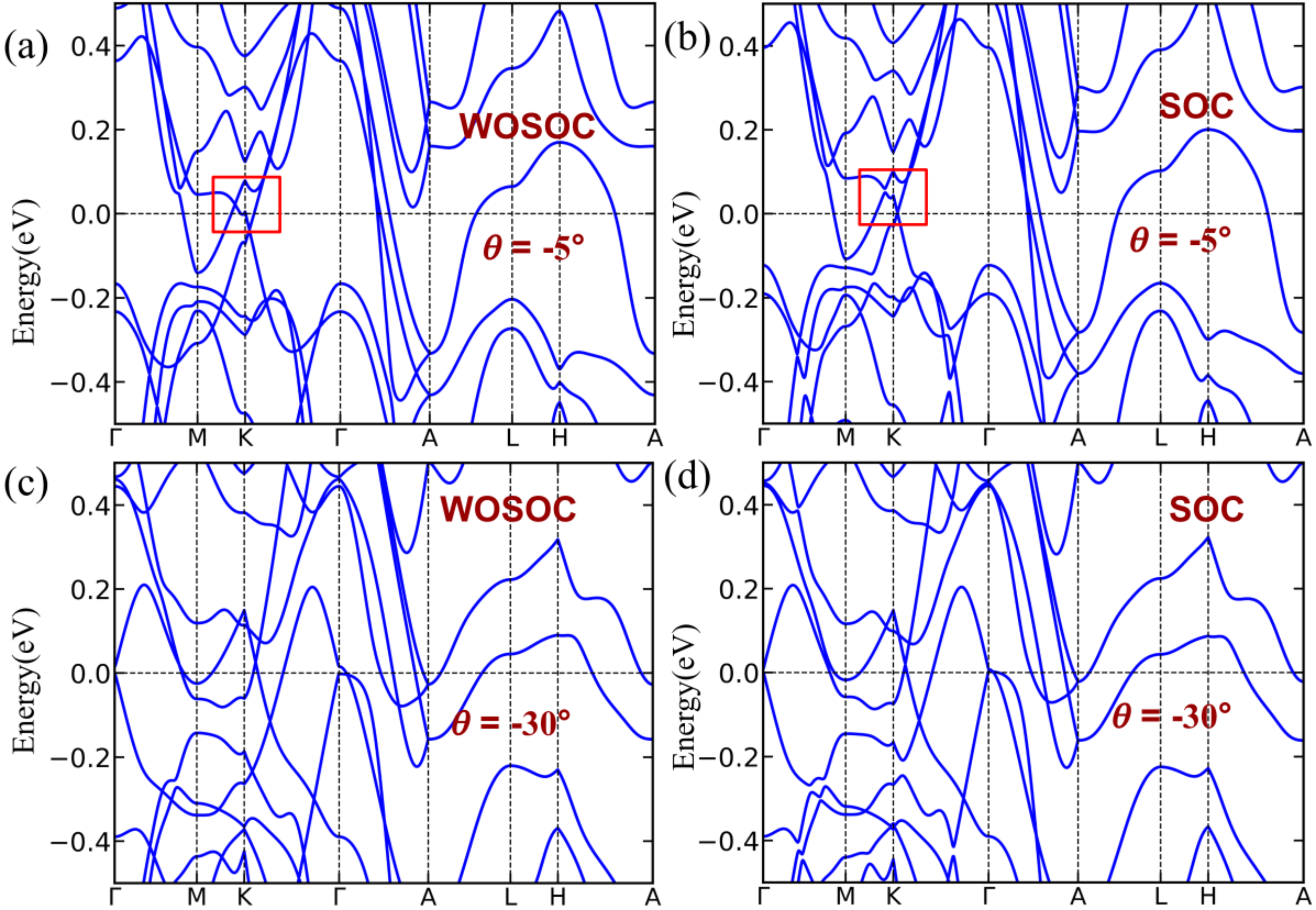}
     \caption{Band structures computed along a high-symmetry path in the Brillouin zone (BZ). Panels (a) and (b) display results for $\theta = -5^\circ$, with (a) showing calculations without spin-orbit coupling (SOC) and (b) with SOC included. Similarly, panels (c) and (d) show the band structures for $\theta = -30^\circ$, again without SOC in (c) and with SOC in (d).} 
     \label{fig:soc_wsoc}
\end{figure}
The ``direct" AFM configuration with $\kappa = +1$ is characterized by the presence of symmetries $\mathcal{M}_x$, $\mathcal{M}_yT$, $\mathcal{M}_zT$, and $C_{3z}$, as illustrated in Fig.~\ref{fig1}(c). However, when a finite \textit{staggered rotation} angle $\theta$ is introduced, the $C_{3z}$ symmetry is broken, while the remaining mirror symmetries and the combination of mirror symmetries with time reversal symmetry, $\mathcal{M}_x$, $\mathcal{M}_yT$, and $\mathcal{M}_zT$, are preserved. This alteration in symmetry raises the need to understand the influence of these mirror symmetries on the Berry curvature in momentum space. The relevant transformations under these symmetries are detailed in the following.
\begin{equation}
    \begin{split}
~ & \mathcal{M}_x: (k_x,k_y,k_z) \rightarrow (-k_x,k_y,k_z) ;~ \Omega^{x}\rightarrow +\Omega^{x} \\
  & \mathcal{M}_yT : (k_x,k_y,k_z) \rightarrow (-k_x,k_y,-k_z) ;~ \Omega^{y}\rightarrow -\Omega^{y} \\
  & \mathcal{M}_zT: (k_x,k_y,k_z) \rightarrow (-k_x,-k_y,k_z) ;~ \Omega^{z}\rightarrow -\Omega^{z} 
    \end{split}
\end{equation}
The components $\Omega^y$ and $\Omega^z$ exhibit antisymmetric behavior under the operations $\mathcal{M}_y T$ and $\mathcal{M}_z T$, respectively. Consequently, the corresponding AHC components, $\sigma_y$ and $\sigma_z$, vanish due to symmetry restrictions. On the other hand, the even nature of $\Omega^x$ under the $\mathcal{M}_x$ operation ensures that only $\sigma_x$ remains nonzero for configurations with a finite \textit{staggered rotation} angle $\theta$. Notably, our \textit{ab initio} calculations reveal that, for all magnetic configurations involving a finite \textit{staggered rotation}, the AHC contribution arises exclusively from $\sigma_{yz}$, which corresponds to the $\sigma_x$ component.

\subsection{Controlling transverse transport through \textit{staggered~rotation}} In this chiral AFM, the \textit{staggered rotation} angle $\theta$ provides an effective means to tune its electronic structure and the resulting anomalous transport properties, such as the AHE and ANE in a Controlled way. We explore in detail how both the AHC and ANC respond to changes in the magnetic configuration. Starting from the ``direct" AFM state, we gradually change $\theta$ within the range of $\pm 15^\circ$. Outside of this range of $\theta$, the band structure undergoes significant alterations. Our focus is on the intrinsic mechanisms of AHC and ANC tied to the material's band structure, particularly the influence of Berry curvature. Our calculations reveal that the transverse conductivity in the $x$-direction ($\sigma_x$) is the only component that remains non-zero for configurations with finite $\theta$, consistent with the symmetry analysis presented earlier (see Fig.~\ref{fig:ahc_tuning}(a) for the variation of AHC as a function of $\theta$).
Our result highlights the pivotal influence of the rotation angle on the control of the AHC. In the starting configuration with $\theta = 0^\circ$ (``direct" AFM state), the AHC is zero. However, even a small increase in $\theta$ causes the AHC to jump to a finite value. Notably, the direction of the \textit{staggered rotation} plays a significant role—positive $\theta$ (counterclockwise) results in a positive AHC, while negative $\theta$ (clockwise) leads to a negative AHC. Moreover, the AHC shows asymmetry for equal rotation angles in opposite directions. The highest AHC, $-310~S\cdot cm^{-1}$, occurs at a $-15^\circ$ rotation, underscoring the substantial impact of this control mechanism. To provide a more intuitive understanding of how the \textit{staggered rotation} angle $\theta$ influences the AHC, Fig.~\ref{fig:AHC_Scan} illustrates the variation of AHC across different energy levels near the Fermi energy for several values of $\theta$, spanning from $-15^\circ$ to $+15^\circ$. These plots distinctly show how the AHC transitions from negative to positive depending on whether the spins are rotated one way or the other (positive or negative $\theta$). The plots also show that the AHC value is strongly influenced by the energy level within the material. These results offer valuable insights into how we can precisely control AHC in this AFM phase of Mn$_3$Sn by adjusting the \textit{staggered~rotation} and electron concentration.\\ 
Let us now turn our attention to the ANE, a captivating electrical phenomenon intrinsically linked to the Berry curvature. This effect emerges in the presence of a temperature gradient, $\nabla T$, within the material, resulting in a measurable response described by the anomalous thermoelectric response tensor~\cite{Guin_2019}. Similar to the behavior of the AHC, Fig.~\ref{fig:ahc_tuning}(a) illustrates how the ANC changes with $\theta$ at a temperature of 300 K. The plot demonstrates the tuning of the ANC component $\alpha_x$ ($\alpha_{yz}$), which shifts from positive to negative, crossing zero as $\theta$ varies from $-15^{\circ}$ to $15^{\circ}$. For the ``direct" AFM configuration at $\theta = 0^{\circ}$, the ANC value is zero, a result of the NR topological band structure. Additionally, the plot highlights that, like the AHC, the ANC exhibits asymmetry depending on the direction of the \textit{staggered rotation} (i.e., the sign of $\theta$). The maximum ANC value is observed at $\theta = -15^{\circ}$, where $\alpha_x = -1.90~\text{A}\cdot\text{m}^{-1}\cdot\text{K}^{-1}$ at T = 300~K. In Fig.~\ref{fig:ANC_Scan}, we present the energy-dependent ANC for various $\theta$ values ranging from $-15^{\circ}$ to $+15^{\circ}$. These plots further confirm the tunability of ANC as a function of $\theta$.\\ \\
Our calculations for the AHC and ANC highlight an important connection: the symmetry-allowed non-zero values of $\sigma_x$ and $\alpha_x$ are highly responsive to the material's internal electronic structure. This sensitivity is particularly evident in their dependence on the chemical potential. Notably, our result demonstrates that both AHC and ANC can be reversed (from positive to negative and vice versa) simply by adjusting the spin rotation angle, $\theta$, in the kagome plane. Interestingly, even a small staggered rotation, which breaks the $C_{3z}$ symmetry, can switch both AHC and ANC from zero to either positive or negative values. This finding provides a powerful method for controlling electrical transport properties in this material.

\section{\label{sec:level4} Microscopic understanding of anomalous transport}
The quantum state depicted in Fig.~\ref{fig:ahc_tuning}(b) as an elliptical nodal ring in momentum space plays a key role in driving unique electrical transport properties when rotational symmetry (C$_{3z}$) is broken due to finite \textit{staggered rotation}. As illustrated in Fig.~\ref{fig:ahc_tuning}(c), a \textit{staggered rotation} ($\theta = -5^\circ$) breaks this symmetry, causing the band crossing points along the elliptical ring to gap out. This symmetry-breaking effect modifies the electronic structure near the Fermi energy, specifically around the high-symmetry `K' point in the hexagonal Brillouin zone (BZ). The formation of the gap in the elliptical nodal ring is attributed to SOC.\\  
To visualize the impact of \textit{staggered rotation} on the material's electronic structure, Fig.~\ref{fig:berry_band}(a)-(f) displays the band structures for different $\theta$ values ranging from $-15^{\circ}$ to $+15^{\circ}$. Alongside each band structure, the Berry curvature component ($\Omega_x$), which influences anomalous transport properties, is plotted. The color bar indicates the magnitude of the Berry curvature. Notably, all non-zero $\theta$ configurations exhibit a SOC-induced band gap near the Fermi energy around the high-symmetry `K' point. This gap results in finite $\Omega_x$ values. A clear trend emerges from the analysis: as $\theta$ increases, the gap between bands (band splitting) also increases. This phenomenon correlates with the growth of in-plane magnetization (not shown here), which is absent at $\theta = 0^\circ$ (the ``direct" AFM configuration). The direction of the in-plane magnetization depends on the rotational sense of the Mn spins, flipping as the sign of $\theta$ changes. Additionally, significant values of $\Omega_x$ are observed along the high-symmetry `MK' direction, coinciding with the nodal line gap location. These SOC-induced gaps, which generate high Berry curvature, are responsible for the intrinsic AHC and ANC. The variation in $\Omega_x$ could be connected to the reversal of magnetization direction when the sign of $\theta$ is inverted. The reversal of the AHC and ANC signs can be attributed to the inversion of the momentum-dependent Berry curvature distribution close to the Fermi energy, as shown within the red box of Fig.~\ref{fig:berry_band}. As $\theta$ changes sign, the contributions of the bands to $\Omega_x$ reverse. For instance, at $\theta = -5^\circ$, the upper band predominantly contributes positive Berry curvature, while the lower band contributes most to negative. At $\theta = +5^\circ$, this pattern flips: the upper band contributes negative Berry curvature, while the lower band contributes positive. This band inversion, which reverses the Berry curvature contributions, causes the AHC to switch from negative to positive and the ANC from positive to negative as the \textit{staggered rotation} angle transitions from negative to positive. This interplay between Berry curvature, SOC-induced band gaps, and magnetization direction underscores the tunable nature of anomalous transport properties in these systems.\\
While the Berry curvature contributions from the gapped band structures provide a clear explanation for the sign change observed in $\sigma_x$ and $\alpha_x$, the dependence of the band positions on $\theta$ appears to be unrelated. This lack of direct correlation may arise from the complex interplay between in-plane magnetization and SOC effects. To simplify the analysis, we focus on the two bands that form the NR in the $\theta = 0^\circ$ configuration. To delve deeper into this phenomenon, Figs.~\ref{fig:BC_2D}(a) and (c) depict the band gap between these two bands in the $k_z = 0$ plane for $\theta = -5^\circ$ and $\theta = +5^\circ$, respectively. The magnitude of the gap varies differently along the NR for opposite signs of $\theta$ with the same magnitude. Consequently, the momentum-dependent Berry curvature distribution differs significantly, as shown in Figs.~\ref{fig:BC_2D}(b) and (d). In contrast to a simple Berry curvature sign flip, the distribution of Berry curvature for $\theta = -5^\circ$ deviates substantially from that of $\theta = +5^\circ$, which may account for the larger values of $\sigma_x$ and $\alpha_x$ observed for negative $\theta$ angles. The induced in-plane magnetization underscores the importance of the absolute sense (positive or negative) of \textit{staggered~rotation} in controlling the transport properties of chiral planar AFMs.\\
Although increasing the staggered rotation angle can modify the band structure, significantly large values ($\theta \gtrsim 20^\circ$) may prove counterproductive. Stronger in-plane magnetization (associated with higher $\theta$) substantially alters bands near the Fermi energy, potentially disrupting the essential band topology (a gapped nodal ring state) required for Berry curvature generation. This reduces the impact of SOC on these bands, thereby restricting the desired anomalous transport effects. Fig.~\ref{fig:soc_wsoc} clearly demonstrate this point. Panels (a) and (b) show the band structure of the $\theta = -5^\circ$ configuration, with and without SOC, respectively. Significantly, SOC opens a small band gap in the nodal ring around the high-symmetry `K' point, as shown within the red box. This gap directly facilitates the substantial SOC-induced anomalous transport observed in this configuration. In sharp contrast, panels (c) and (d) display the band structures for a significantly larger value of $\theta$ ($\theta = -30^\circ$), with and without SOC, respectively. Interestingly, the band crossing point that was observed near the Fermi energy (indicated by red boxes in panels (a) and (b)) is no longer present in this case. This absence aligns with a minimal effect of SOC on the bands close to the Fermi level. Together, these findings indicate that a large staggered rotation interferes with the essential band features needed to further modify the transport phenomenas.\\

\section{\label{sec:level5}Conclusion}
Our \textit{ab initio} calculations and symmetry analysis reveal that a tunable AHC and ANC emerge through \textit{staggered~rotation} within a specific ``direct" ($\kappa = +1$) AFM phase of Mn$_3$Sn. This study demonstrates that even a small \textit{staggered~rotation} within this specific ``direct" 120$^\circ$ AFM arrangement can significantly enhance both AHC and ANC. Moreover, by tuning the magnitude and sign of the \textit{staggered~rotation}, we can precisely control the sign and magnitude of the AHC and ANC. This effect arises from the \textit{staggered~rotation} breaking a specific symmetry ($C_{3z}$ rotational symmetry) that would otherwise cancel out AHC and ANC. In simpler terms, a small \textit{staggered~rotation} unlocks the potential for these intriguing transport properties. This approach offers a promising avenue for experimental realization and potential applications in antiferromagnetic spintronics, where the ability to manipulate AHC and ANC in non-collinear antiferromagnets is crucial.
\begin{acknowledgments}
A.K.N. acknowledges the financial support from Department of Atomic Energy (DAE), Government of India, through the project Basic Research in Physical and Multidisciplinary Sciences via RIN4001. A.K.N. and S.P. acknowledge the computational resources, Kalinga cluster, at National Institute
of Science Education and Research, Bhubaneswar, India. A.K.N. thanks Prof. P. M. Oppeneer for the Swedish
National Infrastructure for Computing (SNIC) facility. A.K.N. and S.P. acknowledge Dr. Charanpreet Singh and Dr. Kush Saha for fruitful discussions. K.S. acknowledges the computing resources provided by the University of Nebraska-Lincoln's Holland Computing Center.
\end{acknowledgments}


\bibliographystyle{apsrev4-2}
\bibliography{references}

\providecommand{\noopsort}[1]{}\providecommand{\singleletter}[1]{#1}%
\begin{thebibliography}{55}%
\makeatletter
\providecommand \@ifxundefined [1]{%
 \@ifx{#1\undefined}
}%
\providecommand \@ifnum [1]{%
 \ifnum #1\expandafter \@firstoftwo
 \else \expandafter \@secondoftwo
 \fi
}%
\providecommand \@ifx [1]{%
 \ifx #1\expandafter \@firstoftwo
 \else \expandafter \@secondoftwo
 \fi
}%
\providecommand \natexlab [1]{#1}%
\providecommand \enquote  [1]{``#1''}%
\providecommand \bibnamefont  [1]{#1}%
\providecommand \bibfnamefont [1]{#1}%
\providecommand \citenamefont [1]{#1}%
\providecommand \href@noop [0]{\@secondoftwo}%
\providecommand \href [0]{\begingroup \@sanitize@url \@href}%
\providecommand \@href[1]{\@@startlink{#1}\@@href}%
\providecommand \@@href[1]{\endgroup#1\@@endlink}%
\providecommand \@sanitize@url [0]{\catcode `\\12\catcode `\$12\catcode
  `\&12\catcode `\#12\catcode `\^12\catcode `\_12\catcode `\%12\relax}%
\providecommand \@@startlink[1]{}%
\providecommand \@@endlink[0]{}%
\providecommand \url  [0]{\begingroup\@sanitize@url \@url }%
\providecommand \@url [1]{\endgroup\@href {#1}{\urlprefix }}%
\providecommand \urlprefix  [0]{URL }%
\providecommand \Eprint [0]{\href }%
\providecommand \doibase [0]{https://doi.org/}%
\providecommand \selectlanguage [0]{\@gobble}%
\providecommand \bibinfo  [0]{\@secondoftwo}%
\providecommand \bibfield  [0]{\@secondoftwo}%
\providecommand \translation [1]{[#1]}%
\providecommand \BibitemOpen [0]{}%
\providecommand \bibitemStop [0]{}%
\providecommand \bibitemNoStop [0]{.\EOS\space}%
\providecommand \EOS [0]{\spacefactor3000\relax}%
\providecommand \BibitemShut  [1]{\csname bibitem#1\endcsname}%
\let\auto@bib@innerbib\@empty
\bibitem [{\citenamefont {Fang}\ \emph {et~al.}(2016)\citenamefont {Fang},
  \citenamefont {Weng}, \citenamefont {Dai},\ and\ \citenamefont
  {Fang}}]{Fang_2016}%
  \BibitemOpen
  \bibfield  {author} {\bibinfo {author} {\bibfnamefont {C.}~\bibnamefont
  {Fang}}, \bibinfo {author} {\bibfnamefont {H.}~\bibnamefont {Weng}}, \bibinfo
  {author} {\bibfnamefont {X.}~\bibnamefont {Dai}},\ and\ \bibinfo {author}
  {\bibfnamefont {Z.}~\bibnamefont {Fang}},\ }\href
  {https://doi.org/10.1088/1674-1056/25/11/117106} {\bibfield  {journal}
  {\bibinfo  {journal} {Chinese Physics B}\ }\textbf {\bibinfo {volume} {25}},\
  \bibinfo {pages} {117106} (\bibinfo {year} {2016})}\BibitemShut {NoStop}%
\bibitem [{\citenamefont {Weng}\ \emph {et~al.}(2015)\citenamefont {Weng},
  \citenamefont {Liang}, \citenamefont {Xu}, \citenamefont {Yu}, \citenamefont
  {Fang}, \citenamefont {Dai},\ and\ \citenamefont
  {Kawazoe}}]{PhysRevB.92.045108}%
  \BibitemOpen
  \bibfield  {author} {\bibinfo {author} {\bibfnamefont {H.}~\bibnamefont
  {Weng}}, \bibinfo {author} {\bibfnamefont {Y.}~\bibnamefont {Liang}},
  \bibinfo {author} {\bibfnamefont {Q.}~\bibnamefont {Xu}}, \bibinfo {author}
  {\bibfnamefont {R.}~\bibnamefont {Yu}}, \bibinfo {author} {\bibfnamefont
  {Z.}~\bibnamefont {Fang}}, \bibinfo {author} {\bibfnamefont {X.}~\bibnamefont
  {Dai}},\ and\ \bibinfo {author} {\bibfnamefont {Y.}~\bibnamefont {Kawazoe}},\
  }\href {https://doi.org/10.1103/PhysRevB.92.045108} {\bibfield  {journal}
  {\bibinfo  {journal} {Phys. Rev. B}\ }\textbf {\bibinfo {volume} {92}},\
  \bibinfo {pages} {045108} (\bibinfo {year} {2015})}\BibitemShut {NoStop}%
\bibitem [{\citenamefont {Bera}\ \emph {et~al.}(2023)\citenamefont {Bera},
  \citenamefont {Chatterjee}, \citenamefont {Pradhan}, \citenamefont {Pradhan},
  \citenamefont {Kalimuddin}, \citenamefont {Bera}, \citenamefont {Nandy},\
  and\ \citenamefont {Mondal}}]{PhysRevB.108.115122}%
  \BibitemOpen
  \bibfield  {author} {\bibinfo {author} {\bibfnamefont {S.}~\bibnamefont
  {Bera}}, \bibinfo {author} {\bibfnamefont {S.}~\bibnamefont {Chatterjee}},
  \bibinfo {author} {\bibfnamefont {S.}~\bibnamefont {Pradhan}}, \bibinfo
  {author} {\bibfnamefont {S.~K.}\ \bibnamefont {Pradhan}}, \bibinfo {author}
  {\bibfnamefont {S.}~\bibnamefont {Kalimuddin}}, \bibinfo {author}
  {\bibfnamefont {A.}~\bibnamefont {Bera}}, \bibinfo {author} {\bibfnamefont
  {A.~K.}\ \bibnamefont {Nandy}},\ and\ \bibinfo {author} {\bibfnamefont
  {M.}~\bibnamefont {Mondal}},\ }\href
  {https://doi.org/10.1103/PhysRevB.108.115122} {\bibfield  {journal} {\bibinfo
   {journal} {Phys. Rev. B}\ }\textbf {\bibinfo {volume} {108}},\ \bibinfo
  {pages} {115122} (\bibinfo {year} {2023})}\BibitemShut {NoStop}%
\bibitem [{\citenamefont {Yan}\ and\ \citenamefont {Felser}(2017)}]{Yan_2017}%
  \BibitemOpen
  \bibfield  {author} {\bibinfo {author} {\bibfnamefont {B.}~\bibnamefont
  {Yan}}\ and\ \bibinfo {author} {\bibfnamefont {C.}~\bibnamefont {Felser}},\
  }\href {https://doi.org/10.1146/annurev-conmatphys-031016-025458} {\bibfield
  {journal} {\bibinfo  {journal} {Annual Review of Condensed Matter Physics}\
  }\textbf {\bibinfo {volume} {8}},\ \bibinfo {pages} {337} (\bibinfo {year}
  {2017})}\BibitemShut {NoStop}%
\bibitem [{\citenamefont {Armitage}\ \emph {et~al.}(2018)\citenamefont
  {Armitage}, \citenamefont {Mele},\ and\ \citenamefont
  {Vishwanath}}]{RevModPhys.90.015001}%
  \BibitemOpen
  \bibfield  {author} {\bibinfo {author} {\bibfnamefont {N.~P.}\ \bibnamefont
  {Armitage}}, \bibinfo {author} {\bibfnamefont {E.~J.}\ \bibnamefont {Mele}},\
  and\ \bibinfo {author} {\bibfnamefont {A.}~\bibnamefont {Vishwanath}},\
  }\href {https://doi.org/10.1103/RevModPhys.90.015001} {\bibfield  {journal}
  {\bibinfo  {journal} {Rev. Mod. Phys.}\ }\textbf {\bibinfo {volume} {90}},\
  \bibinfo {pages} {015001} (\bibinfo {year} {2018})}\BibitemShut {NoStop}%
\bibitem [{\citenamefont {Yang}(2016)}]{Yang_2016}%
  \BibitemOpen
  \bibfield  {author} {\bibinfo {author} {\bibfnamefont {S.~A.}\ \bibnamefont
  {Yang}},\ }\href {https://doi.org/10.1142/s2010324716400038} {\bibfield
  {journal} {\bibinfo  {journal} {{SPIN}}\ }\textbf {\bibinfo {volume} {06}},\
  \bibinfo {pages} {1640003} (\bibinfo {year} {2016})}\BibitemShut {NoStop}%
\bibitem [{\citenamefont {Wan}\ \emph {et~al.}(2011)\citenamefont {Wan},
  \citenamefont {Turner}, \citenamefont {Vishwanath},\ and\ \citenamefont
  {Savrasov}}]{PhysRevB.83.205101}%
  \BibitemOpen
  \bibfield  {author} {\bibinfo {author} {\bibfnamefont {X.}~\bibnamefont
  {Wan}}, \bibinfo {author} {\bibfnamefont {A.~M.}\ \bibnamefont {Turner}},
  \bibinfo {author} {\bibfnamefont {A.}~\bibnamefont {Vishwanath}},\ and\
  \bibinfo {author} {\bibfnamefont {S.~Y.}\ \bibnamefont {Savrasov}},\ }\href
  {https://doi.org/10.1103/PhysRevB.83.205101} {\bibfield  {journal} {\bibinfo
  {journal} {Phys. Rev. B}\ }\textbf {\bibinfo {volume} {83}},\ \bibinfo
  {pages} {205101} (\bibinfo {year} {2011})}\BibitemShut {NoStop}%
\bibitem [{\citenamefont {Li}\ \emph {et~al.}(2020)\citenamefont {Li},
  \citenamefont {Koo}, \citenamefont {Ning}, \citenamefont {Li}, \citenamefont
  {Miao}, \citenamefont {Min}, \citenamefont {Zhu}, \citenamefont {Alem},
  \citenamefont {Liu}, \citenamefont {Mao},\ and\ \citenamefont
  {Yan}}]{article}%
  \BibitemOpen
  \bibfield  {author} {\bibinfo {author} {\bibfnamefont {P.}~\bibnamefont
  {Li}}, \bibinfo {author} {\bibfnamefont {J.}~\bibnamefont {Koo}}, \bibinfo
  {author} {\bibfnamefont {W.}~\bibnamefont {Ning}}, \bibinfo {author}
  {\bibfnamefont {J.}~\bibnamefont {Li}}, \bibinfo {author} {\bibfnamefont
  {L.}~\bibnamefont {Miao}}, \bibinfo {author} {\bibfnamefont {L.}~\bibnamefont
  {Min}}, \bibinfo {author} {\bibfnamefont {Y.}~\bibnamefont {Zhu}}, \bibinfo
  {author} {\bibfnamefont {N.}~\bibnamefont {Alem}}, \bibinfo {author}
  {\bibfnamefont {C.-X.}\ \bibnamefont {Liu}}, \bibinfo {author} {\bibfnamefont
  {Z.}~\bibnamefont {Mao}},\ and\ \bibinfo {author} {\bibfnamefont
  {B.}~\bibnamefont {Yan}},\ }\href
  {https://doi.org/10.1038/s41467-020-17174-9} {\bibfield  {journal} {\bibinfo
  {journal} {Nature Communications}\ }\textbf {\bibinfo {volume} {11}},\
  \bibinfo {pages} {3476} (\bibinfo {year} {2020})}\BibitemShut {NoStop}%
\bibitem [{\citenamefont {Noky}\ \emph {et~al.}(2019)\citenamefont {Noky},
  \citenamefont {Xu}, \citenamefont {Felser},\ and\ \citenamefont
  {Sun}}]{PhysRevB.99.165117}%
  \BibitemOpen
  \bibfield  {author} {\bibinfo {author} {\bibfnamefont {J.}~\bibnamefont
  {Noky}}, \bibinfo {author} {\bibfnamefont {Q.}~\bibnamefont {Xu}}, \bibinfo
  {author} {\bibfnamefont {C.}~\bibnamefont {Felser}},\ and\ \bibinfo {author}
  {\bibfnamefont {Y.}~\bibnamefont {Sun}},\ }\href
  {https://doi.org/https://doi.org/10.1103/PhysRevB.99.165117} {\bibfield
  {journal} {\bibinfo  {journal} {Phys. Rev. B}\ }\textbf {\bibinfo {volume}
  {99}},\ \bibinfo {pages} {165117} (\bibinfo {year} {2019})}\BibitemShut
  {NoStop}%
\bibitem [{\citenamefont {Baltz}\ \emph {et~al.}(2018)\citenamefont {Baltz},
  \citenamefont {Manchon}, \citenamefont {Tsoi}, \citenamefont {Moriyama},
  \citenamefont {Ono},\ and\ \citenamefont
  {Tserkovnyak}}]{RevModPhys.90.015005}%
  \BibitemOpen
  \bibfield  {author} {\bibinfo {author} {\bibfnamefont {V.}~\bibnamefont
  {Baltz}}, \bibinfo {author} {\bibfnamefont {A.}~\bibnamefont {Manchon}},
  \bibinfo {author} {\bibfnamefont {M.}~\bibnamefont {Tsoi}}, \bibinfo {author}
  {\bibfnamefont {T.}~\bibnamefont {Moriyama}}, \bibinfo {author}
  {\bibfnamefont {T.}~\bibnamefont {Ono}},\ and\ \bibinfo {author}
  {\bibfnamefont {Y.}~\bibnamefont {Tserkovnyak}},\ }\href
  {https://doi.org/10.1103/RevModPhys.90.015005} {\bibfield  {journal}
  {\bibinfo  {journal} {Rev. Mod. Phys.}\ }\textbf {\bibinfo {volume} {90}},\
  \bibinfo {pages} {015005} (\bibinfo {year} {2018})}\BibitemShut {NoStop}%
\bibitem [{\citenamefont {Salemi}\ \emph {et~al.}(2019)\citenamefont {Salemi},
  \citenamefont {Berritta}, \citenamefont {Nandy},\ and\ \citenamefont
  {Oppeneer}}]{Salemi_2019}%
  \BibitemOpen
  \bibfield  {author} {\bibinfo {author} {\bibfnamefont {L.}~\bibnamefont
  {Salemi}}, \bibinfo {author} {\bibfnamefont {M.}~\bibnamefont {Berritta}},
  \bibinfo {author} {\bibfnamefont {A.~K.}\ \bibnamefont {Nandy}},\ and\
  \bibinfo {author} {\bibfnamefont {P.~M.}\ \bibnamefont {Oppeneer}},\
  }\bibfield  {journal} {\bibinfo  {journal} {Nature Communications}\ }\textbf
  {\bibinfo {volume} {10}},\ \href {https://doi.org/10.1038/s41467-019-13367-z}
  {10.1038/s41467-019-13367-z} (\bibinfo {year} {2019})\BibitemShut {NoStop}%
\bibitem [{\citenamefont {Jungwirth}\ \emph {et~al.}(2016)\citenamefont
  {Jungwirth}, \citenamefont {Marti}, \citenamefont {Wadley},\ and\
  \citenamefont {Wunderlich}}]{Jungwirth_2016}%
  \BibitemOpen
  \bibfield  {author} {\bibinfo {author} {\bibfnamefont {T.}~\bibnamefont
  {Jungwirth}}, \bibinfo {author} {\bibfnamefont {X.}~\bibnamefont {Marti}},
  \bibinfo {author} {\bibfnamefont {P.}~\bibnamefont {Wadley}},\ and\ \bibinfo
  {author} {\bibfnamefont {J.}~\bibnamefont {Wunderlich}},\ }\href
  {https://doi.org/10.1038/nnano.2016.18} {\bibfield  {journal} {\bibinfo
  {journal} {Nature Nanotechnology}\ }\textbf {\bibinfo {volume} {11}},\
  \bibinfo {pages} {231–241} (\bibinfo {year} {2016})}\BibitemShut {NoStop}%
\bibitem [{\citenamefont {Feng}\ \emph {et~al.}(2015)\citenamefont {Feng},
  \citenamefont {Guo}, \citenamefont {Zhou}, \citenamefont {Yao},\ and\
  \citenamefont {Niu}}]{PhysRevB.92.144426}%
  \BibitemOpen
  \bibfield  {author} {\bibinfo {author} {\bibfnamefont {W.}~\bibnamefont
  {Feng}}, \bibinfo {author} {\bibfnamefont {G.-Y.}\ \bibnamefont {Guo}},
  \bibinfo {author} {\bibfnamefont {J.}~\bibnamefont {Zhou}}, \bibinfo {author}
  {\bibfnamefont {Y.}~\bibnamefont {Yao}},\ and\ \bibinfo {author}
  {\bibfnamefont {Q.}~\bibnamefont {Niu}},\ }\href
  {https://doi.org/10.1103/PhysRevB.92.144426} {\bibfield  {journal} {\bibinfo
  {journal} {Phys. Rev. B}\ }\textbf {\bibinfo {volume} {92}},\ \bibinfo
  {pages} {144426} (\bibinfo {year} {2015})}\BibitemShut {NoStop}%
\bibitem [{\citenamefont {Chen}\ \emph {et~al.}(2014)\citenamefont {Chen},
  \citenamefont {Niu},\ and\ \citenamefont
  {MacDonald}}]{PhysRevLett.112.017205}%
  \BibitemOpen
  \bibfield  {author} {\bibinfo {author} {\bibfnamefont {H.}~\bibnamefont
  {Chen}}, \bibinfo {author} {\bibfnamefont {Q.}~\bibnamefont {Niu}},\ and\
  \bibinfo {author} {\bibfnamefont {A.~H.}\ \bibnamefont {MacDonald}},\ }\href
  {https://doi.org/10.1103/PhysRevLett.112.017205} {\bibfield  {journal}
  {\bibinfo  {journal} {Phys. Rev. Lett.}\ }\textbf {\bibinfo {volume} {112}},\
  \bibinfo {pages} {017205} (\bibinfo {year} {2014})}\BibitemShut {NoStop}%
\bibitem [{\citenamefont {Kübler}\ and\ \citenamefont
  {Felser}(2017)}]{Kubler_2017}%
  \BibitemOpen
  \bibfield  {author} {\bibinfo {author} {\bibfnamefont {J.}~\bibnamefont
  {Kübler}}\ and\ \bibinfo {author} {\bibfnamefont {C.}~\bibnamefont
  {Felser}},\ }\href {https://doi.org/10.1209/0295-5075/120/47002} {\bibfield
  {journal} {\bibinfo  {journal} {EPL (Europhysics Letters)}\ }\textbf
  {\bibinfo {volume} {120}},\ \bibinfo {pages} {47002} (\bibinfo {year}
  {2017})}\BibitemShut {NoStop}%
\bibitem [{\citenamefont {Yang}\ \emph {et~al.}(2017)\citenamefont {Yang},
  \citenamefont {Sun}, \citenamefont {Zhang}, \citenamefont {Shi},
  \citenamefont {Parkin},\ and\ \citenamefont {Yan}}]{Yang_2017}%
  \BibitemOpen
  \bibfield  {author} {\bibinfo {author} {\bibfnamefont {H.}~\bibnamefont
  {Yang}}, \bibinfo {author} {\bibfnamefont {Y.}~\bibnamefont {Sun}}, \bibinfo
  {author} {\bibfnamefont {Y.}~\bibnamefont {Zhang}}, \bibinfo {author}
  {\bibfnamefont {W.-J.}\ \bibnamefont {Shi}}, \bibinfo {author} {\bibfnamefont
  {S.~S.~P.}\ \bibnamefont {Parkin}},\ and\ \bibinfo {author} {\bibfnamefont
  {B.}~\bibnamefont {Yan}},\ }\href {https://doi.org/10.1088/1367-2630/aa5487}
  {\bibfield  {journal} {\bibinfo  {journal} {New Journal of Physics}\ }\textbf
  {\bibinfo {volume} {19}},\ \bibinfo {pages} {015008} (\bibinfo {year}
  {2017})}\BibitemShut {NoStop}%
\bibitem [{\citenamefont {Wang}\ \emph {et~al.}(2022)\citenamefont {Wang},
  \citenamefont {Yan}, \citenamefont {Zhou}, \citenamefont {Chen},
  \citenamefont {Feng}, \citenamefont {Qin}, \citenamefont {Meng},
  \citenamefont {Liu},\ and\ \citenamefont {Liu}}]{WANG2022100878}%
  \BibitemOpen
  \bibfield  {author} {\bibinfo {author} {\bibfnamefont {X.}~\bibnamefont
  {Wang}}, \bibinfo {author} {\bibfnamefont {H.}~\bibnamefont {Yan}}, \bibinfo
  {author} {\bibfnamefont {X.}~\bibnamefont {Zhou}}, \bibinfo {author}
  {\bibfnamefont {H.}~\bibnamefont {Chen}}, \bibinfo {author} {\bibfnamefont
  {Z.}~\bibnamefont {Feng}}, \bibinfo {author} {\bibfnamefont {P.}~\bibnamefont
  {Qin}}, \bibinfo {author} {\bibfnamefont {Z.}~\bibnamefont {Meng}}, \bibinfo
  {author} {\bibfnamefont {L.}~\bibnamefont {Liu}},\ and\ \bibinfo {author}
  {\bibfnamefont {Z.}~\bibnamefont {Liu}},\ }\href
  {https://doi.org/https://doi.org/10.1016/j.mtphys.2022.100878} {\bibfield
  {journal} {\bibinfo  {journal} {Materials Today Physics}\ }\textbf {\bibinfo
  {volume} {28}},\ \bibinfo {pages} {100878} (\bibinfo {year}
  {2022})}\BibitemShut {NoStop}%
\bibitem [{\citenamefont {Park}\ \emph {et~al.}(2018)\citenamefont {Park},
  \citenamefont {Oh}, \citenamefont {Uhlířová}, \citenamefont {Jackson},
  \citenamefont {Deák}, \citenamefont {Szunyogh}, \citenamefont {Lee},
  \citenamefont {Cho}, \citenamefont {Kim}, \citenamefont {Walker},
  \citenamefont {Adroja}, \citenamefont {Sechovský},\ and\ \citenamefont
  {Park}}]{Park_2018}%
  \BibitemOpen
  \bibfield  {author} {\bibinfo {author} {\bibfnamefont {P.}~\bibnamefont
  {Park}}, \bibinfo {author} {\bibfnamefont {J.}~\bibnamefont {Oh}}, \bibinfo
  {author} {\bibfnamefont {K.}~\bibnamefont {Uhlířová}}, \bibinfo {author}
  {\bibfnamefont {J.}~\bibnamefont {Jackson}}, \bibinfo {author} {\bibfnamefont
  {A.}~\bibnamefont {Deák}}, \bibinfo {author} {\bibfnamefont
  {L.}~\bibnamefont {Szunyogh}}, \bibinfo {author} {\bibfnamefont {K.~H.}\
  \bibnamefont {Lee}}, \bibinfo {author} {\bibfnamefont {H.}~\bibnamefont
  {Cho}}, \bibinfo {author} {\bibfnamefont {H.-L.}\ \bibnamefont {Kim}},
  \bibinfo {author} {\bibfnamefont {H.~C.}\ \bibnamefont {Walker}}, \bibinfo
  {author} {\bibfnamefont {D.}~\bibnamefont {Adroja}}, \bibinfo {author}
  {\bibfnamefont {V.}~\bibnamefont {Sechovský}},\ and\ \bibinfo {author}
  {\bibfnamefont {J.-G.}\ \bibnamefont {Park}},\ }\bibfield  {journal}
  {\bibinfo  {journal} {npj Quantum Materials}\ }\textbf {\bibinfo {volume}
  {3}},\ \href {https://doi.org/10.1038/s41535-018-0137-9}
  {10.1038/s41535-018-0137-9} (\bibinfo {year} {2018})\BibitemShut {NoStop}%
\bibitem [{\citenamefont {Singh}\ \emph {et~al.}(2024)\citenamefont {Singh},
  \citenamefont {Jamaluddin}, \citenamefont {Pradhan}, \citenamefont {Nandy},
  \citenamefont {Tokunaga}, \citenamefont {Avdeev},\ and\ \citenamefont
  {Nayak}}]{Singh_2024}%
  \BibitemOpen
  \bibfield  {author} {\bibinfo {author} {\bibfnamefont {C.}~\bibnamefont
  {Singh}}, \bibinfo {author} {\bibfnamefont {S.}~\bibnamefont {Jamaluddin}},
  \bibinfo {author} {\bibfnamefont {S.}~\bibnamefont {Pradhan}}, \bibinfo
  {author} {\bibfnamefont {A.~K.}\ \bibnamefont {Nandy}}, \bibinfo {author}
  {\bibfnamefont {M.}~\bibnamefont {Tokunaga}}, \bibinfo {author}
  {\bibfnamefont {M.}~\bibnamefont {Avdeev}},\ and\ \bibinfo {author}
  {\bibfnamefont {A.~K.}\ \bibnamefont {Nayak}},\ }\bibfield  {journal}
  {\bibinfo  {journal} {npj Quantum Materials}\ }\textbf {\bibinfo {volume}
  {9}},\ \href {https://doi.org/10.1038/s41535-024-00657-z}
  {10.1038/s41535-024-00657-z} (\bibinfo {year} {2024})\BibitemShut {NoStop}%
\bibitem [{\citenamefont {Bhattacharya}\ \emph {et~al.}(2024)\citenamefont
  {Bhattacharya}, \citenamefont {Bharatwaj}, \citenamefont {Singh},
  \citenamefont {Gupta}, \citenamefont {Khasanov}, \citenamefont {Kanungo},
  \citenamefont {Nayak},\ and\ \citenamefont {Majumder}}]{PhysRevB.110.094432}%
  \BibitemOpen
  \bibfield  {author} {\bibinfo {author} {\bibfnamefont {K.}~\bibnamefont
  {Bhattacharya}}, \bibinfo {author} {\bibfnamefont {A.~K.}\ \bibnamefont
  {Bharatwaj}}, \bibinfo {author} {\bibfnamefont {C.}~\bibnamefont {Singh}},
  \bibinfo {author} {\bibfnamefont {R.}~\bibnamefont {Gupta}}, \bibinfo
  {author} {\bibfnamefont {R.}~\bibnamefont {Khasanov}}, \bibinfo {author}
  {\bibfnamefont {S.}~\bibnamefont {Kanungo}}, \bibinfo {author} {\bibfnamefont
  {A.~K.}\ \bibnamefont {Nayak}},\ and\ \bibinfo {author} {\bibfnamefont
  {M.}~\bibnamefont {Majumder}},\ }\href
  {https://doi.org/10.1103/PhysRevB.110.094432} {\bibfield  {journal} {\bibinfo
   {journal} {Phys. Rev. B}\ }\textbf {\bibinfo {volume} {110}},\ \bibinfo
  {pages} {094432} (\bibinfo {year} {2024})}\BibitemShut {NoStop}%
\bibitem [{\citenamefont {Kübler}\ and\ \citenamefont
  {Felser}(2014)}]{K_bler_2014}%
  \BibitemOpen
  \bibfield  {author} {\bibinfo {author} {\bibfnamefont {J.}~\bibnamefont
  {Kübler}}\ and\ \bibinfo {author} {\bibfnamefont {C.}~\bibnamefont
  {Felser}},\ }\href {https://doi.org/10.1209/0295-5075/108/67001} {\bibfield
  {journal} {\bibinfo  {journal} {{EPL} (Europhysics Letters)}\ }\textbf
  {\bibinfo {volume} {108}},\ \bibinfo {pages} {67001} (\bibinfo {year}
  {2014})}\BibitemShut {NoStop}%
\bibitem [{\citenamefont {Nakatsuji}\ \emph {et~al.}(2015)\citenamefont
  {Nakatsuji}, \citenamefont {Kiyohara},\ and\ \citenamefont
  {Higo}}]{nakatsuji2015large}%
  \BibitemOpen
  \bibfield  {author} {\bibinfo {author} {\bibfnamefont {S.}~\bibnamefont
  {Nakatsuji}}, \bibinfo {author} {\bibfnamefont {N.}~\bibnamefont
  {Kiyohara}},\ and\ \bibinfo {author} {\bibfnamefont {T.}~\bibnamefont
  {Higo}},\ }\href {https://doi.org/10.1038/nature15723} {\bibfield  {journal}
  {\bibinfo  {journal} {Nature}\ }\textbf {\bibinfo {volume} {527}},\ \bibinfo
  {pages} {212} (\bibinfo {year} {2015})}\BibitemShut {NoStop}%
\bibitem [{\citenamefont {Nayak}\ \emph {et~al.}(2016)\citenamefont {Nayak},
  \citenamefont {Fischer}, \citenamefont {Sun}, \citenamefont {Yan},
  \citenamefont {Karel}, \citenamefont {Komarek}, \citenamefont {Shekhar},
  \citenamefont {Kumar}, \citenamefont {Schnelle}, \citenamefont {Kübler},
  \citenamefont {Felser},\ and\ \citenamefont
  {Parkin}}]{doi:10.1126/sciadv.1501870}%
  \BibitemOpen
  \bibfield  {author} {\bibinfo {author} {\bibfnamefont {A.~K.}\ \bibnamefont
  {Nayak}}, \bibinfo {author} {\bibfnamefont {J.~E.}\ \bibnamefont {Fischer}},
  \bibinfo {author} {\bibfnamefont {Y.}~\bibnamefont {Sun}}, \bibinfo {author}
  {\bibfnamefont {B.}~\bibnamefont {Yan}}, \bibinfo {author} {\bibfnamefont
  {J.}~\bibnamefont {Karel}}, \bibinfo {author} {\bibfnamefont {A.~C.}\
  \bibnamefont {Komarek}}, \bibinfo {author} {\bibfnamefont {C.}~\bibnamefont
  {Shekhar}}, \bibinfo {author} {\bibfnamefont {N.}~\bibnamefont {Kumar}},
  \bibinfo {author} {\bibfnamefont {W.}~\bibnamefont {Schnelle}}, \bibinfo
  {author} {\bibfnamefont {J.}~\bibnamefont {Kübler}}, \bibinfo {author}
  {\bibfnamefont {C.}~\bibnamefont {Felser}},\ and\ \bibinfo {author}
  {\bibfnamefont {S.~S.~P.}\ \bibnamefont {Parkin}},\ }\href
  {https://doi.org/10.1126/sciadv.1501870} {\bibfield  {journal} {\bibinfo
  {journal} {Science Advances}\ }\textbf {\bibinfo {volume} {2}},\ \bibinfo
  {pages} {e1501870} (\bibinfo {year} {2016})}\BibitemShut {NoStop}%
\bibitem [{\citenamefont {Zhang}\ \emph {et~al.}(2017)\citenamefont {Zhang},
  \citenamefont {Sun}, \citenamefont {Yang}, \citenamefont
  {\ifmmode~\check{Z}\else \v{Z}\fi{}elezn\'y}, \citenamefont {Parkin},
  \citenamefont {Felser},\ and\ \citenamefont {Yan}}]{PhysRevB.95.075128}%
  \BibitemOpen
  \bibfield  {author} {\bibinfo {author} {\bibfnamefont {Y.}~\bibnamefont
  {Zhang}}, \bibinfo {author} {\bibfnamefont {Y.}~\bibnamefont {Sun}}, \bibinfo
  {author} {\bibfnamefont {H.}~\bibnamefont {Yang}}, \bibinfo {author}
  {\bibfnamefont {J.}~\bibnamefont {\ifmmode~\check{Z}\else
  \v{Z}\fi{}elezn\'y}}, \bibinfo {author} {\bibfnamefont {S.~P.~P.}\
  \bibnamefont {Parkin}}, \bibinfo {author} {\bibfnamefont {C.}~\bibnamefont
  {Felser}},\ and\ \bibinfo {author} {\bibfnamefont {B.}~\bibnamefont {Yan}},\
  }\href {https://doi.org/10.1103/PhysRevB.95.075128} {\bibfield  {journal}
  {\bibinfo  {journal} {Phys. Rev. B}\ }\textbf {\bibinfo {volume} {95}},\
  \bibinfo {pages} {075128} (\bibinfo {year} {2017})}\BibitemShut {NoStop}%
\bibitem [{\citenamefont {Ikhlas}\ \emph {et~al.}(2017)\citenamefont {Ikhlas},
  \citenamefont {Tomita}, \citenamefont {Koretsune}, \citenamefont {Suzuki},
  \citenamefont {Nishio-Hamane}, \citenamefont {Arita}, \citenamefont {Otani},\
  and\ \citenamefont {Nakatsuji}}]{Ikhlas_2017}%
  \BibitemOpen
  \bibfield  {author} {\bibinfo {author} {\bibfnamefont {M.}~\bibnamefont
  {Ikhlas}}, \bibinfo {author} {\bibfnamefont {T.}~\bibnamefont {Tomita}},
  \bibinfo {author} {\bibfnamefont {T.}~\bibnamefont {Koretsune}}, \bibinfo
  {author} {\bibfnamefont {M.-T.}\ \bibnamefont {Suzuki}}, \bibinfo {author}
  {\bibfnamefont {D.}~\bibnamefont {Nishio-Hamane}}, \bibinfo {author}
  {\bibfnamefont {R.}~\bibnamefont {Arita}}, \bibinfo {author} {\bibfnamefont
  {Y.}~\bibnamefont {Otani}},\ and\ \bibinfo {author} {\bibfnamefont
  {S.}~\bibnamefont {Nakatsuji}},\ }\href {https://doi.org/10.1038/nphys4181}
  {\bibfield  {journal} {\bibinfo  {journal} {Nature Physics}\ }\textbf
  {\bibinfo {volume} {13}},\ \bibinfo {pages} {1085} (\bibinfo {year}
  {2017})}\BibitemShut {NoStop}%
\bibitem [{\citenamefont {Li}\ \emph {et~al.}(2017)\citenamefont {Li},
  \citenamefont {Xu}, \citenamefont {Ding}, \citenamefont {Wang}, \citenamefont
  {Shen}, \citenamefont {Lu}, \citenamefont {Zhu},\ and\ \citenamefont
  {Behnia}}]{PhysRevLett.119.056601}%
  \BibitemOpen
  \bibfield  {author} {\bibinfo {author} {\bibfnamefont {X.}~\bibnamefont
  {Li}}, \bibinfo {author} {\bibfnamefont {L.}~\bibnamefont {Xu}}, \bibinfo
  {author} {\bibfnamefont {L.}~\bibnamefont {Ding}}, \bibinfo {author}
  {\bibfnamefont {J.}~\bibnamefont {Wang}}, \bibinfo {author} {\bibfnamefont
  {M.}~\bibnamefont {Shen}}, \bibinfo {author} {\bibfnamefont {X.}~\bibnamefont
  {Lu}}, \bibinfo {author} {\bibfnamefont {Z.}~\bibnamefont {Zhu}},\ and\
  \bibinfo {author} {\bibfnamefont {K.}~\bibnamefont {Behnia}},\ }\href
  {https://doi.org/10.1103/PhysRevLett.119.056601} {\bibfield  {journal}
  {\bibinfo  {journal} {Phys. Rev. Lett.}\ }\textbf {\bibinfo {volume} {119}},\
  \bibinfo {pages} {056601} (\bibinfo {year} {2017})}\BibitemShut {NoStop}%
\bibitem [{\citenamefont {Sharma}\ \emph {et~al.}(2023)\citenamefont {Sharma},
  \citenamefont {Nepal},\ and\ \citenamefont {Budhani}}]{PhysRevB.108.144435}%
  \BibitemOpen
  \bibfield  {author} {\bibinfo {author} {\bibfnamefont {V.}~\bibnamefont
  {Sharma}}, \bibinfo {author} {\bibfnamefont {R.}~\bibnamefont {Nepal}},\ and\
  \bibinfo {author} {\bibfnamefont {R.~C.}\ \bibnamefont {Budhani}},\ }\href
  {https://doi.org/10.1103/PhysRevB.108.144435} {\bibfield  {journal} {\bibinfo
   {journal} {Phys. Rev. B}\ }\textbf {\bibinfo {volume} {108}},\ \bibinfo
  {pages} {144435} (\bibinfo {year} {2023})}\BibitemShut {NoStop}%
\bibitem [{\citenamefont {Dong}\ \emph {et~al.}(2022)\citenamefont {Dong},
  \citenamefont {Li}, \citenamefont {Gurung}, \citenamefont {Zhu},
  \citenamefont {Zhang}, \citenamefont {Zheng}, \citenamefont {Tsymbal},\ and\
  \citenamefont {Zhang}}]{PhysRevLett.128.197201}%
  \BibitemOpen
  \bibfield  {author} {\bibinfo {author} {\bibfnamefont {J.}~\bibnamefont
  {Dong}}, \bibinfo {author} {\bibfnamefont {X.}~\bibnamefont {Li}}, \bibinfo
  {author} {\bibfnamefont {G.}~\bibnamefont {Gurung}}, \bibinfo {author}
  {\bibfnamefont {M.}~\bibnamefont {Zhu}}, \bibinfo {author} {\bibfnamefont
  {P.}~\bibnamefont {Zhang}}, \bibinfo {author} {\bibfnamefont
  {F.}~\bibnamefont {Zheng}}, \bibinfo {author} {\bibfnamefont {E.~Y.}\
  \bibnamefont {Tsymbal}},\ and\ \bibinfo {author} {\bibfnamefont
  {J.}~\bibnamefont {Zhang}},\ }\href
  {https://doi.org/10.1103/PhysRevLett.128.197201} {\bibfield  {journal}
  {\bibinfo  {journal} {Phys. Rev. Lett.}\ }\textbf {\bibinfo {volume} {128}},\
  \bibinfo {pages} {197201} (\bibinfo {year} {2022})}\BibitemShut {NoStop}%
\bibitem [{\citenamefont {Yu}\ \emph {et~al.}(2021)\citenamefont {Yu},
  \citenamefont {Wu}, \citenamefont {He}, \citenamefont {Guo}, \citenamefont
  {Fang}, \citenamefont {Zhang}, \citenamefont {Wong}, \citenamefont {Xu},
  \citenamefont {Han},\ and\ \citenamefont {Wang}}]{10.1063/5.0045627}%
  \BibitemOpen
  \bibfield  {author} {\bibinfo {author} {\bibfnamefont {T.}~\bibnamefont
  {Yu}}, \bibinfo {author} {\bibfnamefont {H.}~\bibnamefont {Wu}}, \bibinfo
  {author} {\bibfnamefont {H.}~\bibnamefont {He}}, \bibinfo {author}
  {\bibfnamefont {C.}~\bibnamefont {Guo}}, \bibinfo {author} {\bibfnamefont
  {C.}~\bibnamefont {Fang}}, \bibinfo {author} {\bibfnamefont {P.}~\bibnamefont
  {Zhang}}, \bibinfo {author} {\bibfnamefont {K.~L.}\ \bibnamefont {Wong}},
  \bibinfo {author} {\bibfnamefont {S.}~\bibnamefont {Xu}}, \bibinfo {author}
  {\bibfnamefont {X.}~\bibnamefont {Han}},\ and\ \bibinfo {author}
  {\bibfnamefont {K.~L.}\ \bibnamefont {Wang}},\ }\href
  {https://doi.org/10.1063/5.0045627} {\bibfield  {journal} {\bibinfo
  {journal} {APL Materials}\ }\textbf {\bibinfo {volume} {9}},\ \bibinfo
  {pages} {041111} (\bibinfo {year} {2021})},\ \Eprint
  {https://arxiv.org/abs/https://pubs.aip.org/aip/apm/article-pdf/doi/10.1063/5.0045627/14566673/041111\_1\_online.pdf}
  {https://pubs.aip.org/aip/apm/article-pdf/doi/10.1063/5.0045627/14566673/041111\_1\_online.pdf}
  \BibitemShut {NoStop}%
\bibitem [{\citenamefont {Kawamura}(2001)}]{Kawamura_2001}%
  \BibitemOpen
  \bibfield  {author} {\bibinfo {author} {\bibfnamefont {H.}~\bibnamefont
  {Kawamura}},\ }\href {https://doi.org/10.1139/p01-111} {\bibfield  {journal}
  {\bibinfo  {journal} {Canadian Journal of Physics}\ }\textbf {\bibinfo
  {volume} {79}},\ \bibinfo {pages} {1447} (\bibinfo {year}
  {2001})}\BibitemShut {NoStop}%
\bibitem [{\citenamefont {Pradhan}\ \emph {et~al.}(2023)\citenamefont
  {Pradhan}, \citenamefont {Samanta}, \citenamefont {Saha},\ and\ \citenamefont
  {Nandy}}]{Pradhan_2023}%
  \BibitemOpen
  \bibfield  {author} {\bibinfo {author} {\bibfnamefont {S.}~\bibnamefont
  {Pradhan}}, \bibinfo {author} {\bibfnamefont {K.}~\bibnamefont {Samanta}},
  \bibinfo {author} {\bibfnamefont {K.}~\bibnamefont {Saha}},\ and\ \bibinfo
  {author} {\bibfnamefont {A.~K.}\ \bibnamefont {Nandy}},\ }\bibfield
  {journal} {\bibinfo  {journal} {Communications Physics}\ }\textbf {\bibinfo
  {volume} {6}},\ \href {https://doi.org/10.1038/s42005-023-01385-9}
  {10.1038/s42005-023-01385-9} (\bibinfo {year} {2023})\BibitemShut {NoStop}%
\bibitem [{\citenamefont {Chen}\ \emph {et~al.}(2021)\citenamefont {Chen},
  \citenamefont {Tomita}, \citenamefont {Minami}, \citenamefont {Fu},
  \citenamefont {Koretsune}, \citenamefont {Kitatani}, \citenamefont
  {Muhammad}, \citenamefont {Nishio-Hamane}, \citenamefont {Ishii},
  \citenamefont {Ishii} \emph {et~al.}}]{chen2021anomalous}%
  \BibitemOpen
  \bibfield  {author} {\bibinfo {author} {\bibfnamefont {T.}~\bibnamefont
  {Chen}}, \bibinfo {author} {\bibfnamefont {T.}~\bibnamefont {Tomita}},
  \bibinfo {author} {\bibfnamefont {S.}~\bibnamefont {Minami}}, \bibinfo
  {author} {\bibfnamefont {M.}~\bibnamefont {Fu}}, \bibinfo {author}
  {\bibfnamefont {T.}~\bibnamefont {Koretsune}}, \bibinfo {author}
  {\bibfnamefont {M.}~\bibnamefont {Kitatani}}, \bibinfo {author}
  {\bibfnamefont {I.}~\bibnamefont {Muhammad}}, \bibinfo {author}
  {\bibfnamefont {D.}~\bibnamefont {Nishio-Hamane}}, \bibinfo {author}
  {\bibfnamefont {R.}~\bibnamefont {Ishii}}, \bibinfo {author} {\bibfnamefont
  {F.}~\bibnamefont {Ishii}}, \emph {et~al.},\ }\href
  {https://doi.org/10.1038/s41467-020-20838-1} {\bibfield  {journal} {\bibinfo
  {journal} {Nature communications}\ }\textbf {\bibinfo {volume} {12}},\
  \bibinfo {pages} {1} (\bibinfo {year} {2021})}\BibitemShut {NoStop}%
\bibitem [{\citenamefont {Singh}\ \emph {et~al.}(2020)\citenamefont {Singh},
  \citenamefont {Singh}, \citenamefont {Pradhan}, \citenamefont {Srihari},
  \citenamefont {Poswal}, \citenamefont {Nath}, \citenamefont {Nandy},\ and\
  \citenamefont {Nayak}}]{PhysRevResearch.2.043366}%
  \BibitemOpen
  \bibfield  {author} {\bibinfo {author} {\bibfnamefont {C.}~\bibnamefont
  {Singh}}, \bibinfo {author} {\bibfnamefont {V.}~\bibnamefont {Singh}},
  \bibinfo {author} {\bibfnamefont {G.}~\bibnamefont {Pradhan}}, \bibinfo
  {author} {\bibfnamefont {V.}~\bibnamefont {Srihari}}, \bibinfo {author}
  {\bibfnamefont {H.~K.}\ \bibnamefont {Poswal}}, \bibinfo {author}
  {\bibfnamefont {R.}~\bibnamefont {Nath}}, \bibinfo {author} {\bibfnamefont
  {A.~K.}\ \bibnamefont {Nandy}},\ and\ \bibinfo {author} {\bibfnamefont
  {A.~K.}\ \bibnamefont {Nayak}},\ }\href
  {https://doi.org/10.1103/PhysRevResearch.2.043366} {\bibfield  {journal}
  {\bibinfo  {journal} {Phys. Rev. Research}\ }\textbf {\bibinfo {volume}
  {2}},\ \bibinfo {pages} {043366} (\bibinfo {year} {2020})}\BibitemShut
  {NoStop}%
\bibitem [{\citenamefont {Zhang}\ \emph {et~al.}(2021)\citenamefont {Zhang},
  \citenamefont {Feng}, \citenamefont {Xu}, \citenamefont {Hao},\ and\
  \citenamefont {Du}}]{temperature}%
  \BibitemOpen
  \bibfield  {author} {\bibinfo {author} {\bibfnamefont {H.}~\bibnamefont
  {Zhang}}, \bibinfo {author} {\bibfnamefont {H.}~\bibnamefont {Feng}},
  \bibinfo {author} {\bibfnamefont {X.}~\bibnamefont {Xu}}, \bibinfo {author}
  {\bibfnamefont {W.}~\bibnamefont {Hao}},\ and\ \bibinfo {author}
  {\bibfnamefont {Y.}~\bibnamefont {Du}},\ }\href
  {https://doi.org/https://doi.org/10.1002/qute.202100073} {\bibfield
  {journal} {\bibinfo  {journal} {Advanced Quantum Technologies}\ }\textbf
  {\bibinfo {volume} {4}},\ \bibinfo {pages} {2100073} (\bibinfo {year}
  {2021})}\BibitemShut {NoStop}%
\bibitem [{\citenamefont {Ikhlas}\ \emph {et~al.}(2022)\citenamefont {Ikhlas},
  \citenamefont {Dasgupta}, \citenamefont {Theuss}, \citenamefont {Higo},
  \citenamefont {Kittaka}, \citenamefont {Ramshaw}, \citenamefont
  {Tchernyshyov}, \citenamefont {Hicks},\ and\ \citenamefont
  {Nakatsuji}}]{Pizzo}%
  \BibitemOpen
  \bibfield  {author} {\bibinfo {author} {\bibfnamefont {M.}~\bibnamefont
  {Ikhlas}}, \bibinfo {author} {\bibfnamefont {S.}~\bibnamefont {Dasgupta}},
  \bibinfo {author} {\bibfnamefont {F.}~\bibnamefont {Theuss}}, \bibinfo
  {author} {\bibfnamefont {T.}~\bibnamefont {Higo}}, \bibinfo {author}
  {\bibfnamefont {S.}~\bibnamefont {Kittaka}}, \bibinfo {author} {\bibfnamefont
  {B.~J.}\ \bibnamefont {Ramshaw}}, \bibinfo {author} {\bibfnamefont
  {O.}~\bibnamefont {Tchernyshyov}}, \bibinfo {author} {\bibfnamefont {C.~W.}\
  \bibnamefont {Hicks}},\ and\ \bibinfo {author} {\bibfnamefont
  {S.}~\bibnamefont {Nakatsuji}},\ }\href
  {https://doi.org/10.1038/s41567-022-01645-5} {\bibfield  {journal} {\bibinfo
  {journal} {Nature Physics}\ }\textbf {\bibinfo {volume} {18}},\ \bibinfo
  {pages} {1086–1093} (\bibinfo {year} {2022})}\BibitemShut {NoStop}%
\bibitem [{\citenamefont {Miwa}\ \emph {et~al.}(2021)\citenamefont {Miwa},
  \citenamefont {Iihama}, \citenamefont {Nomoto}, \citenamefont {Tomita},
  \citenamefont {Higo}, \citenamefont {Ikhlas}, \citenamefont {Sakamoto},
  \citenamefont {Otani}, \citenamefont {Mizukami}, \citenamefont {Arita},\ and\
  \citenamefont {Nakatsuji}}]{Miwa}%
  \BibitemOpen
  \bibfield  {author} {\bibinfo {author} {\bibfnamefont {S.}~\bibnamefont
  {Miwa}}, \bibinfo {author} {\bibfnamefont {S.}~\bibnamefont {Iihama}},
  \bibinfo {author} {\bibfnamefont {T.}~\bibnamefont {Nomoto}}, \bibinfo
  {author} {\bibfnamefont {T.}~\bibnamefont {Tomita}}, \bibinfo {author}
  {\bibfnamefont {T.}~\bibnamefont {Higo}}, \bibinfo {author} {\bibfnamefont
  {M.}~\bibnamefont {Ikhlas}}, \bibinfo {author} {\bibfnamefont
  {S.}~\bibnamefont {Sakamoto}}, \bibinfo {author} {\bibfnamefont
  {Y.}~\bibnamefont {Otani}}, \bibinfo {author} {\bibfnamefont
  {S.}~\bibnamefont {Mizukami}}, \bibinfo {author} {\bibfnamefont
  {R.}~\bibnamefont {Arita}},\ and\ \bibinfo {author} {\bibfnamefont
  {S.}~\bibnamefont {Nakatsuji}},\ }\bibfield  {journal} {\bibinfo  {journal}
  {Small Science}\ }\textbf {\bibinfo {volume} {1}},\ \href
  {https://doi.org/10.1002/smsc.202000062} {10.1002/smsc.202000062} (\bibinfo
  {year} {2021})\BibitemShut {NoStop}%
\bibitem [{\citenamefont {Li}\ \emph {et~al.}(2022)\citenamefont {Li},
  \citenamefont {Jiang}, \citenamefont {Meng}, \citenamefont {Zuo},
  \citenamefont {Zhu}, \citenamefont {Balents},\ and\ \citenamefont
  {Behnia}}]{PhysRevB.106.L020402}%
  \BibitemOpen
  \bibfield  {author} {\bibinfo {author} {\bibfnamefont {X.}~\bibnamefont
  {Li}}, \bibinfo {author} {\bibfnamefont {S.}~\bibnamefont {Jiang}}, \bibinfo
  {author} {\bibfnamefont {Q.}~\bibnamefont {Meng}}, \bibinfo {author}
  {\bibfnamefont {H.}~\bibnamefont {Zuo}}, \bibinfo {author} {\bibfnamefont
  {Z.}~\bibnamefont {Zhu}}, \bibinfo {author} {\bibfnamefont {L.}~\bibnamefont
  {Balents}},\ and\ \bibinfo {author} {\bibfnamefont {K.}~\bibnamefont
  {Behnia}},\ }\href {https://doi.org/10.1103/PhysRevB.106.L020402} {\bibfield
  {journal} {\bibinfo  {journal} {Phys. Rev. B}\ }\textbf {\bibinfo {volume}
  {106}},\ \bibinfo {pages} {L020402} (\bibinfo {year} {2022})}\BibitemShut
  {NoStop}%
\bibitem [{fle()}]{fleur}%
  \BibitemOpen
  \href {https://www.flapw.de/MaX-6.0/} {\bibinfo {title}
  {{www.flapw.de}}}\BibitemShut {NoStop}%
\bibitem [{\citenamefont {Kresse}\ and\ \citenamefont {Joubert}(1999)}]{vasp}%
  \BibitemOpen
  \bibfield  {author} {\bibinfo {author} {\bibfnamefont {G.}~\bibnamefont
  {Kresse}}\ and\ \bibinfo {author} {\bibfnamefont {D.}~\bibnamefont
  {Joubert}},\ }\href {https://doi.org/10.1103/PhysRevB.59.1758} {\bibfield
  {journal} {\bibinfo  {journal} {Phys. Rev. B}\ }\textbf {\bibinfo {volume}
  {59}},\ \bibinfo {pages} {1758} (\bibinfo {year} {1999})}\BibitemShut
  {NoStop}%
\bibitem [{\citenamefont {Perdew}\ \emph {et~al.}(1996)\citenamefont {Perdew},
  \citenamefont {Burke},\ and\ \citenamefont {Ernzerhof}}]{pbe}%
  \BibitemOpen
  \bibfield  {author} {\bibinfo {author} {\bibfnamefont {J.~P.}\ \bibnamefont
  {Perdew}}, \bibinfo {author} {\bibfnamefont {K.}~\bibnamefont {Burke}},\ and\
  \bibinfo {author} {\bibfnamefont {M.}~\bibnamefont {Ernzerhof}},\ }\href
  {https://doi.org/10.1103/PhysRevLett.77.3865} {\bibfield  {journal} {\bibinfo
   {journal} {Phys. Rev. Lett.}\ }\textbf {\bibinfo {volume} {77}},\ \bibinfo
  {pages} {3865} (\bibinfo {year} {1996})}\BibitemShut {NoStop}%
\bibitem [{\citenamefont {Bl\"ochl}(1994)}]{paw2}%
  \BibitemOpen
  \bibfield  {author} {\bibinfo {author} {\bibfnamefont {P.~E.}\ \bibnamefont
  {Bl\"ochl}},\ }\href {https://doi.org/10.1103/PhysRevB.50.17953} {\bibfield
  {journal} {\bibinfo  {journal} {Phys. Rev. B}\ }\textbf {\bibinfo {volume}
  {50}},\ \bibinfo {pages} {17953} (\bibinfo {year} {1994})}\BibitemShut
  {NoStop}%
\bibitem [{\citenamefont {Monkhorst}\ and\ \citenamefont {Pack}(1976)}]{monk}%
  \BibitemOpen
  \bibfield  {author} {\bibinfo {author} {\bibfnamefont {H.~J.}\ \bibnamefont
  {Monkhorst}}\ and\ \bibinfo {author} {\bibfnamefont {J.~D.}\ \bibnamefont
  {Pack}},\ }\href {https://doi.org/10.1103/PhysRevB.13.5188} {\bibfield
  {journal} {\bibinfo  {journal} {Phys. Rev. B}\ }\textbf {\bibinfo {volume}
  {13}},\ \bibinfo {pages} {5188} (\bibinfo {year} {1976})}\BibitemShut
  {NoStop}%
\bibitem [{\citenamefont {{Vosko}}\ \emph {et~al.}(1980)\citenamefont
  {{Vosko}}, \citenamefont {{Wilk}},\ and\ \citenamefont
  {{Nusair}}}]{1980CaJPh..58.1200V}%
  \BibitemOpen
  \bibfield  {author} {\bibinfo {author} {\bibfnamefont {S.~H.}\ \bibnamefont
  {{Vosko}}}, \bibinfo {author} {\bibfnamefont {L.}~\bibnamefont {{Wilk}}},\
  and\ \bibinfo {author} {\bibfnamefont {M.}~\bibnamefont {{Nusair}}},\ }\href
  {https://doi.org/10.1139/p80-159} {\bibfield  {journal} {\bibinfo  {journal}
  {Canadian Journal of Physics}\ }\textbf {\bibinfo {volume} {59}},\ \bibinfo
  {pages} {1200} (\bibinfo {year} {1980})}\BibitemShut {NoStop}%
\bibitem [{\citenamefont {Marzari}\ \emph {et~al.}(2012)\citenamefont
  {Marzari}, \citenamefont {Mostofi}, \citenamefont {Yates}, \citenamefont
  {Souza},\ and\ \citenamefont {Vanderbilt}}]{wan2}%
  \BibitemOpen
  \bibfield  {author} {\bibinfo {author} {\bibfnamefont {N.}~\bibnamefont
  {Marzari}}, \bibinfo {author} {\bibfnamefont {A.~A.}\ \bibnamefont
  {Mostofi}}, \bibinfo {author} {\bibfnamefont {J.~R.}\ \bibnamefont {Yates}},
  \bibinfo {author} {\bibfnamefont {I.}~\bibnamefont {Souza}},\ and\ \bibinfo
  {author} {\bibfnamefont {D.}~\bibnamefont {Vanderbilt}},\ }\href
  {https://doi.org/10.1103/RevModPhys.84.1419} {\bibfield  {journal} {\bibinfo
  {journal} {Rev. Mod. Phys.}\ }\textbf {\bibinfo {volume} {84}},\ \bibinfo
  {pages} {1419} (\bibinfo {year} {2012})}\BibitemShut {NoStop}%
\bibitem [{\citenamefont {Mostofi}\ \emph {et~al.}(2014)\citenamefont
  {Mostofi}, \citenamefont {Yates}, \citenamefont {Pizzi}, \citenamefont {Lee},
  \citenamefont {Souza}, \citenamefont {Vanderbilt},\ and\ \citenamefont
  {Marzari}}]{wan4}%
  \BibitemOpen
  \bibfield  {author} {\bibinfo {author} {\bibfnamefont {A.~A.}\ \bibnamefont
  {Mostofi}}, \bibinfo {author} {\bibfnamefont {J.~R.}\ \bibnamefont {Yates}},
  \bibinfo {author} {\bibfnamefont {G.}~\bibnamefont {Pizzi}}, \bibinfo
  {author} {\bibfnamefont {Y.-S.}\ \bibnamefont {Lee}}, \bibinfo {author}
  {\bibfnamefont {I.}~\bibnamefont {Souza}}, \bibinfo {author} {\bibfnamefont
  {D.}~\bibnamefont {Vanderbilt}},\ and\ \bibinfo {author} {\bibfnamefont
  {N.}~\bibnamefont {Marzari}},\ }\href
  {https://doi.org/https://doi.org/10.1016/j.cpc.2014.05.003} {\bibfield
  {journal} {\bibinfo  {journal} {Computer Physics Communications}\ }\textbf
  {\bibinfo {volume} {185}},\ \bibinfo {pages} {2309} (\bibinfo {year}
  {2014})}\BibitemShut {NoStop}%
\bibitem [{\citenamefont {Wang}\ \emph {et~al.}(2006)\citenamefont {Wang},
  \citenamefont {Yates}, \citenamefont {Souza},\ and\ \citenamefont
  {Vanderbilt}}]{wan1}%
  \BibitemOpen
  \bibfield  {author} {\bibinfo {author} {\bibfnamefont {X.}~\bibnamefont
  {Wang}}, \bibinfo {author} {\bibfnamefont {J.~R.}\ \bibnamefont {Yates}},
  \bibinfo {author} {\bibfnamefont {I.}~\bibnamefont {Souza}},\ and\ \bibinfo
  {author} {\bibfnamefont {D.}~\bibnamefont {Vanderbilt}},\ }\href
  {https://doi.org/10.1103/PhysRevB.74.195118} {\bibfield  {journal} {\bibinfo
  {journal} {Phys. Rev. B}\ }\textbf {\bibinfo {volume} {74}},\ \bibinfo
  {pages} {195118} (\bibinfo {year} {2006})}\BibitemShut {NoStop}%
\bibitem [{\citenamefont {Freimuth}\ \emph {et~al.}(2008)\citenamefont
  {Freimuth}, \citenamefont {Mokrousov}, \citenamefont {Wortmann},
  \citenamefont {Heinze},\ and\ \citenamefont {Bl\"ugel}}]{Freimuth-2008}%
  \BibitemOpen
  \bibfield  {author} {\bibinfo {author} {\bibfnamefont {F.}~\bibnamefont
  {Freimuth}}, \bibinfo {author} {\bibfnamefont {Y.}~\bibnamefont {Mokrousov}},
  \bibinfo {author} {\bibfnamefont {D.}~\bibnamefont {Wortmann}}, \bibinfo
  {author} {\bibfnamefont {S.}~\bibnamefont {Heinze}},\ and\ \bibinfo {author}
  {\bibfnamefont {S.}~\bibnamefont {Bl\"ugel}},\ }\href
  {https://doi.org/10.1103/PhysRevB.78.035120} {\bibfield  {journal} {\bibinfo
  {journal} {Phys. Rev. B}\ }\textbf {\bibinfo {volume} {78}},\ \bibinfo
  {pages} {035120} (\bibinfo {year} {2008})}\BibitemShut {NoStop}%
\bibitem [{\citenamefont {Yao}\ \emph {et~al.}(2004)\citenamefont {Yao},
  \citenamefont {Kleinman}, \citenamefont {MacDonald}, \citenamefont {Sinova},
  \citenamefont {Jungwirth}, \citenamefont {Wang}, \citenamefont {Wang},\ and\
  \citenamefont {Niu}}]{Yao-2004}%
  \BibitemOpen
  \bibfield  {author} {\bibinfo {author} {\bibfnamefont {Y.}~\bibnamefont
  {Yao}}, \bibinfo {author} {\bibfnamefont {L.}~\bibnamefont {Kleinman}},
  \bibinfo {author} {\bibfnamefont {A.~H.}\ \bibnamefont {MacDonald}}, \bibinfo
  {author} {\bibfnamefont {J.}~\bibnamefont {Sinova}}, \bibinfo {author}
  {\bibfnamefont {T.}~\bibnamefont {Jungwirth}}, \bibinfo {author}
  {\bibfnamefont {D.-s.}\ \bibnamefont {Wang}}, \bibinfo {author}
  {\bibfnamefont {E.}~\bibnamefont {Wang}},\ and\ \bibinfo {author}
  {\bibfnamefont {Q.}~\bibnamefont {Niu}},\ }\href
  {https://doi.org/10.1103/PhysRevLett.92.037204} {\bibfield  {journal}
  {\bibinfo  {journal} {Phys. Rev. Lett.}\ }\textbf {\bibinfo {volume} {92}},\
  \bibinfo {pages} {037204} (\bibinfo {year} {2004})}\BibitemShut {NoStop}%
\bibitem [{\citenamefont {Xiao}\ \emph {et~al.}(2006)\citenamefont {Xiao},
  \citenamefont {Yao}, \citenamefont {Fang},\ and\ \citenamefont
  {Niu}}]{PhysRevLett.97.026603}%
  \BibitemOpen
  \bibfield  {author} {\bibinfo {author} {\bibfnamefont {D.}~\bibnamefont
  {Xiao}}, \bibinfo {author} {\bibfnamefont {Y.}~\bibnamefont {Yao}}, \bibinfo
  {author} {\bibfnamefont {Z.}~\bibnamefont {Fang}},\ and\ \bibinfo {author}
  {\bibfnamefont {Q.}~\bibnamefont {Niu}},\ }\href
  {https://doi.org/10.1103/PhysRevLett.97.026603} {\bibfield  {journal}
  {\bibinfo  {journal} {Phys. Rev. Lett.}\ }\textbf {\bibinfo {volume} {97}},\
  \bibinfo {pages} {026603} (\bibinfo {year} {2006})}\BibitemShut {NoStop}%
\bibitem [{\citenamefont {Xiao}\ \emph {et~al.}(2010)\citenamefont {Xiao},
  \citenamefont {Chang},\ and\ \citenamefont {Niu}}]{RevModPhys.82.1959}%
  \BibitemOpen
  \bibfield  {author} {\bibinfo {author} {\bibfnamefont {D.}~\bibnamefont
  {Xiao}}, \bibinfo {author} {\bibfnamefont {M.-C.}\ \bibnamefont {Chang}},\
  and\ \bibinfo {author} {\bibfnamefont {Q.}~\bibnamefont {Niu}},\ }\href
  {https://doi.org/10.1103/RevModPhys.82.1959} {\bibfield  {journal} {\bibinfo
  {journal} {Rev. Mod. Phys.}\ }\textbf {\bibinfo {volume} {82}},\ \bibinfo
  {pages} {1959} (\bibinfo {year} {2010})}\BibitemShut {NoStop}%
\bibitem [{\citenamefont {Tsirkin}(2021)}]{Tsirkin_2021}%
  \BibitemOpen
  \bibfield  {author} {\bibinfo {author} {\bibfnamefont {S.~S.}\ \bibnamefont
  {Tsirkin}},\ }\bibfield  {journal} {\bibinfo  {journal} {npj Computational
  Materials}\ }\textbf {\bibinfo {volume} {7}},\ \href
  {https://doi.org/10.1038/s41524-021-00498-5} {10.1038/s41524-021-00498-5}
  (\bibinfo {year} {2021})\BibitemShut {NoStop}%
\bibitem [{\citenamefont {Nagaosa}\ \emph {et~al.}(2010)\citenamefont
  {Nagaosa}, \citenamefont {Sinova}, \citenamefont {Onoda}, \citenamefont
  {MacDonald},\ and\ \citenamefont {Ong}}]{RevModPhys.82.1539}%
  \BibitemOpen
  \bibfield  {author} {\bibinfo {author} {\bibfnamefont {N.}~\bibnamefont
  {Nagaosa}}, \bibinfo {author} {\bibfnamefont {J.}~\bibnamefont {Sinova}},
  \bibinfo {author} {\bibfnamefont {S.}~\bibnamefont {Onoda}}, \bibinfo
  {author} {\bibfnamefont {A.~H.}\ \bibnamefont {MacDonald}},\ and\ \bibinfo
  {author} {\bibfnamefont {N.~P.}\ \bibnamefont {Ong}},\ }\href
  {https://doi.org/10.1103/RevModPhys.82.1539} {\bibfield  {journal} {\bibinfo
  {journal} {Rev. Mod. Phys.}\ }\textbf {\bibinfo {volume} {82}},\ \bibinfo
  {pages} {1539} (\bibinfo {year} {2010})}\BibitemShut {NoStop}%
\bibitem [{\citenamefont {Le}\ \emph {et~al.}(2021)\citenamefont {Le},
  \citenamefont {Felser},\ and\ \citenamefont {Sun}}]{PhysRevB.104.125145}%
  \BibitemOpen
  \bibfield  {author} {\bibinfo {author} {\bibfnamefont {C.}~\bibnamefont
  {Le}}, \bibinfo {author} {\bibfnamefont {C.}~\bibnamefont {Felser}},\ and\
  \bibinfo {author} {\bibfnamefont {Y.}~\bibnamefont {Sun}},\ }\href
  {https://doi.org/10.1103/PhysRevB.104.125145} {\bibfield  {journal} {\bibinfo
   {journal} {Phys. Rev. B}\ }\textbf {\bibinfo {volume} {104}},\ \bibinfo
  {pages} {125145} (\bibinfo {year} {2021})}\BibitemShut {NoStop}%
\bibitem [{\citenamefont {Suzuki}\ \emph {et~al.}(2017)\citenamefont {Suzuki},
  \citenamefont {Koretsune}, \citenamefont {Ochi},\ and\ \citenamefont
  {Arita}}]{PhysRevB.95.094406}%
  \BibitemOpen
  \bibfield  {author} {\bibinfo {author} {\bibfnamefont {M.-T.}\ \bibnamefont
  {Suzuki}}, \bibinfo {author} {\bibfnamefont {T.}~\bibnamefont {Koretsune}},
  \bibinfo {author} {\bibfnamefont {M.}~\bibnamefont {Ochi}},\ and\ \bibinfo
  {author} {\bibfnamefont {R.}~\bibnamefont {Arita}},\ }\href
  {https://doi.org/10.1103/PhysRevB.95.094406} {\bibfield  {journal} {\bibinfo
  {journal} {Phys. Rev. B}\ }\textbf {\bibinfo {volume} {95}},\ \bibinfo
  {pages} {094406} (\bibinfo {year} {2017})}\BibitemShut {NoStop}%
\bibitem [{\citenamefont {Guin}\ \emph {et~al.}(2019)\citenamefont {Guin},
  \citenamefont {Manna}, \citenamefont {Noky}, \citenamefont {Watzman},
  \citenamefont {Fu}, \citenamefont {Kumar}, \citenamefont {Schnelle},
  \citenamefont {Shekhar}, \citenamefont {Sun}, \citenamefont {Gooth},\ and\
  \citenamefont {Felser}}]{Guin_2019}%
  \BibitemOpen
  \bibfield  {author} {\bibinfo {author} {\bibfnamefont {S.~N.}\ \bibnamefont
  {Guin}}, \bibinfo {author} {\bibfnamefont {K.}~\bibnamefont {Manna}},
  \bibinfo {author} {\bibfnamefont {J.}~\bibnamefont {Noky}}, \bibinfo {author}
  {\bibfnamefont {S.~J.}\ \bibnamefont {Watzman}}, \bibinfo {author}
  {\bibfnamefont {C.}~\bibnamefont {Fu}}, \bibinfo {author} {\bibfnamefont
  {N.}~\bibnamefont {Kumar}}, \bibinfo {author} {\bibfnamefont
  {W.}~\bibnamefont {Schnelle}}, \bibinfo {author} {\bibfnamefont
  {C.}~\bibnamefont {Shekhar}}, \bibinfo {author} {\bibfnamefont
  {Y.}~\bibnamefont {Sun}}, \bibinfo {author} {\bibfnamefont {J.}~\bibnamefont
  {Gooth}},\ and\ \bibinfo {author} {\bibfnamefont {C.}~\bibnamefont
  {Felser}},\ }\bibfield  {journal} {\bibinfo  {journal} {NPG Asia Materials}\
  }\textbf {\bibinfo {volume} {11}},\ \href
  {https://doi.org/10.1038/s41427-019-0116-z} {10.1038/s41427-019-0116-z}
  (\bibinfo {year} {2019})\BibitemShut {NoStop}%
\end{thebibliography}%
\end{document}